\documentclass[preprint,5p,number,times,sort, compress]{elsarticle}

\usepackage{lineno}
\usepackage[colorlinks,
            linkcolor=blue, 
            anchorcolor=blue,
            citecolor=blue, 
            ]{hyperref}

\usepackage{color}
\usepackage{amsmath}
\modulolinenumbers[1]
\usepackage[utf8]{inputenc}
\usepackage{graphicx}
\usepackage{subcaption}
\usepackage{amsmath}
\usepackage{booktabs}

\usepackage{amssymb}
\usepackage{ulem}
\usepackage{multirow}
\usepackage{soul}
\setstcolor{red}

\usepackage{makecell}

\journal{EPJC}

\begin{document}

\begin{frontmatter}

\title{PMT Waveform Simulation and Reconstruction with Conditional Diffusion Network}

\author[firstaddress]{Kainan Liu\texorpdfstring{{}}}
\author[firstaddress]{Jingyu Huang}
\cortext[mail]{Corresponding author}
\author[firstaddress]{\texorpdfstring{{}Guihong Huang \corref{mail}}{}}
\ead{huanggh@wyu.edu.cn}
\author[firstaddress]{Jianyi Luo}

\address[firstaddress]{Wuyi University, Jiangmen 529020, China}

\begin{abstract}
Photomultiplier tubes (PMTs) are widely employed in particle and nuclear physics experiments. The accuracy of PMT waveform reconstruction directly impacts the detector's spatial and energy resolution. A key challenge arises when multiple photons arrive within a few nanoseconds, making it difficult to resolve individual photoelectrons (PEs). Although supervised deep learning methods have surpassed traditional methods in performance, their practical applicability is limited by the lack of ground-truth PE labels in real data. To address this issue, we propose an innovative weakly supervised waveform simulation and reconstruction approach based on a bidirectional conditional diffusion network framework. The method is fully data-driven and requires only raw waveforms and coarse estimates of PE information as input. It first employs a PE-conditioned diffusion model to simulate realistic waveforms from PE sequences, thereby learning the features of overlapping waveforms. Subsequently, these simulated waveforms are used to train a waveform-conditioned diffusion model to reconstruct the PE sequences from waveforms, reinforcing the learning of features of overlapping waveforms. Through iterative refinement between the two conditional diffusion processes, the model progressively improves reconstruction accuracy. Experimental results demonstrate that the proposed method achieves 99\% of the normalized PE-number resolution averaged over 1-5 p.e. and 80\% of the timing resolution attained by fully supervised learning. 

\end{abstract}


\end{frontmatter}

\section{Introduction}
Recent years have witnessed breakthroughs in neutrino experiments. The Daya Bay experiment precisely measured the mixing angle $\theta_{13}$\cite{DayaBay}, while T2K\cite{T2K} and NOvA\cite{NOvA} reported strong evidence for CP violation in the lepton sector. Additionally, IceCube\cite{IceCube} and KM3NeT/ARCA\cite{KM3NeT} have successively detected ultra-high-energy astrophysical neutrinos.   

The liquid scintillator (LS)-based Jiangmen Underground Neutrino Observatory (JUNO)\cite{JUNO} aims to resolve the neutrino mass ordering by studying reactor antineutrino oscillations. Meanwhile, water-based experiments such as Hyper-Kamiokande \cite{HyperK} and KM3NeT/ORCA\cite{ORCA} target the neutrino mass ordering and the CP-violating phase, through accelerator or atmospheric neutrino oscillation analyses. Notably, JUNO demands unprecedented energy resolution, whereas Hyper-K requires high precision in both energy and direction reconstruction. 

In both LS and water Cherenkov detectors, photomultiplier tubes (PMTs) serve as the primary photon sensors, detecting faint Cherenkov or scintillation light. Since the number of detected photons is directly correlated with event energy\cite{PEMLE}, and their arrival times are tightly linked to the event vertex\cite{QTMLE}, precise photon counting and timing from PMT waveform are critical for accurate energy and vertex reconstruction. 

Traditional high-accuracy waveform reconstruction methods primarily rely on waveform fitting and deconvolution. Waveform fitting techniques infer the underlying signal by matching observed waveforms to expected templates, for instance, the Lawson–Hanson non-negative least squares method used in IceCube\cite{LawsHans} and the FSMP-based approach proposed in Ref.~\cite{FSMP}. Deconvolution methods, on the other hand, recover the original signal by suppressing noise and inverting the detector response function, such as the Gaussian-filter-based deconvolution employed by Daya Bay\cite{GaussFilt} and the short-time Fourier transform technique described in Ref.~\cite{ShortTime}. Compared to simple thresholding, these methods significantly improve photoelectron (PE) counting and timing resolution in low-photon regime. However, they heavily depend on accurate prior knowledge of the detector response, and their performance degrades markedly when resolving multi-PE separated by only a few nanoseconds. 

Recently, machine learning techniques, particularly those excelling in complex pattern recognition, have been increasingly adopted in particle physics. Studies show that supervised convolutional neural networks (CNNs) can substantially enhance PE counting accuracy, thereby improving JUNO's energy resolution\cite{MLPhoton}. Nevertheless, supervised approaches require ground-truth PE labels, which are generally inaccessible in real data. In contrast, generative models such as Generative Adversarial Networks (GANs) and diffusion models have demonstrated remarkable capability in simulating high-quality detector responses in high-energy physics\cite{GANATLAS, KamLSim, DFpc, DFeps}, offering a promising avenue for generating realistic training data. 

In this work, we propose, for the first time, a bidirectional conditional diffusion network framework for synergistic waveform simulation and reconstruction. The framework comprises two components: a PE-conditioned diffusion model that simulates waveforms from PE sequences, and a waveform-cond
itioned diffusion model that reconstructs PE sequences from waveforms. 
%Importantly, our method not only enhances reconstruction accuracy but also yields smooth, high-resolution single-photoelectron charge spectra. 
Although the PE conditions contain errors, the network achieves effective weakly supervised learning through careful training sample selection and iterative refinement. As a benchmark, we also evaluate the performance of both diffusion models and ResNet50 under fully supervised learning. 

The paper is organized as follows: Section II describes the statistical properties of the dataset; Section III details the network architecture; Section IV presents the ideal performance under supervised learning; Section V introduces the weakly supervised learning framework and analyzes its performance; and Section VI concludes the study.

\section{Monte Carlo Simulation of PMT Waveforms} 

This study employs waveform data generated by electronics Monte Carlo (EMC) to evaluate and optimize the simulation quality of the PE-conditioned diffusion model and the reconstruction accuracy of the waveform-conditioned diffusion model. To accurately reproduce the characteristics of PMT waveforms in LS detectors, the EMC incorporates key physical parameters inspired by the JUNO experiment: the LS fluorescence time distribution, single-photoelectron (sPE) pulse shape, sPE charge spectrum, and electronic noise level. These ingredients ensure that the simulated waveforms reflect real detector responses. Furthermore, the distribution of number of PE (nPE) in the EMC is carefully designed to match realistic experimental conditions. This section describes the simulation parameters and the resulting dataset used for model training. 

\subsection{Electronics Monte Carlo Simulation} 

In a PMT, incident photons strike the photocathode and produce PEs. These PEs are amplified under high voltage and digitized by a fast analog-to-digital converter (FADC), yielding an output voltage waveform. The EMC constructs waveforms by modeling the following components: nPE distribution, sPE time distribution, pulse shape, charge spectrum, and noise spectrum. The procedure consists of four steps: 

\begin{itemize}
    \item \textbf{PEs generation}: The number of PEs and their arrival times are sampled from prescribed distributions. These sampled values constitute the Monte Carlo truth. 
\item \textbf{Pulse convolution}: Each PE is convolved with a sPE pulse template that includes a main peak, an overshoot, and multiple reflection peaks. The amplitude of each sPE follows a Gaussian distribution corresponding to 0.3 p.e. charge resolution.
\item \textbf{Noise and baseline addition}: Electronic noise is modeled as white noise with a standard deviation of 0.7 mV. No baseline offset is applied.
\item \textbf{Digitization}: The analog waveform is digitized at a sampling rate of 1 GHz with 10-bit ADC precision, using standard rounding.

\end{itemize} 

\begin{figure}
    \centering
    \includegraphics[width=1\linewidth]{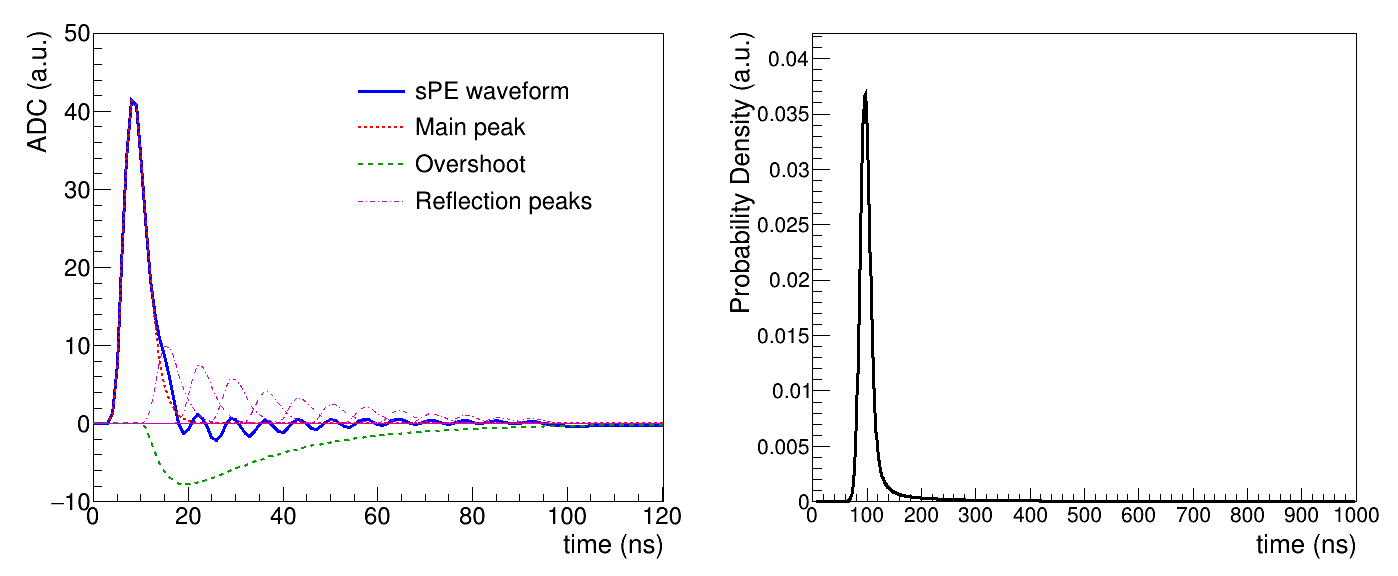}
    \caption{
    Left: The averaged sPE waveform comprises three components: the main peak, overshoot, and reflection peaks. 
    Right: Probability density function (PDF) of the LS fluorescence time, obtained by convolving a four-exponential scintillation decay model with the instrument response function.
    }
    \label{fig:Single_p.e.Wave-FluoescenceTime}
\end{figure}

 The sPE pulse shape used in this work is based on PMT measurements from the Daya Bay experiment\cite{FlashADC}, as illustrated in Fig.\ref{fig:Single_p.e.Wave-FluoescenceTime}. The following subsection details the nPE and time distributions adopted in this study. 

\subsection{nPE and Time Distributions}
\label{sec:nPEDist} 
PEs detected by the PMT originate primarily from LS light and dark noise. LS photons are produced when an incident particle deposits energy in the LS, and the resulting total count of detected photons is positively correlated with the deposited energy. In contrast, dark-noise PEs arise from spontaneous  emission in the PMT and are uncorrelated with the energy of incident particle. 

In this study, we assume that the nPE from LS $k^{\mathrm{L}}$ and the nPE from dark noise $k^{\mathrm{D}}$ follow independent Poisson distributions:
$P(k^{\mathrm{L}}, \mu^{\mathrm{L}})$ and $P(k^{\mathrm{D}},\mu^{\mathrm{D}})$, where $\mu^{\mathrm{L}}$ and $\mu^{\mathrm{D}}$ denote the mean nPE from LS and dark noise, respectively. For MeV-scale neutrino events near the center of the JUNO detector, $\mu^{\mathrm{L}}$ is on the order of 0.1 p.e. Assuming a PMT dark-count rate of 30 kHz and a waveform readout window of 1000 ns, the corresponding $\mu^{\mathrm{D}}$ is 0.03 p.e. Clearly, increasing $k^{\mathrm{L}}$ or $k^{\mathrm{D}}$ results in more severe waveform overlap. 

Most neutrino experiments, including JUNO\cite{CalibJUNO} and SNO\cite{CalibSNO}, are equipped with tunable laser systems and gamma sources of various energies. By combining data with different $\mu^{\mathrm{L}}$, one can obtain waveforms whose nPE approximate a uniform distribution in practice. 

Following Ref.\cite{FluorTime}, we adopt a four-exponential model for the distribution of LS fluorescence time. To simulate the time response of microchannel plate PMTs (MCP-PMTs) in JUNO, the transit-time spread is set to $\sigma_{\mathrm{PMT}}$=8 ns\cite{junoLPMT}, as shown in Fig.~\ref{fig:Single_p.e.Wave-FluoescenceTime}. The arrival times of dark-noise PEs are modeled as uniformly distributed over the readout window. Evidently, a narrower time distribution increases the degree of waveform overlap. 

To investigate the impact of nPE and time distributions on the performance of diffusion model, we define three representative scenarios based on realistic experimental conditions: 

\begin{itemize}
    \item \textbf{UT-UPE}: Hit time of PEs follows a uniform distribution, and the nPE is uniformly distributed between 0 and 10 p.e. This configuration features large time spread, broad PE-multiplicity, and low degree of waveform overlap, representing an idealized experimental condition. 
    \item \textbf{LT-xPE}: Hit time of LS PEs follows the LS time distribution, and nPE of LS PEs follows a Poisson distribution with mean $x$ p.e., supplemented by dark-noise PEs (uniform in time, Poisson-distributed with mean 0.03 p.e.). This scenario exhibits narrow time spread, limited PE-multiplicity, and moderate degree of waveform overlap, corresponding to typical physical events.
    \item \textbf{LT-UPE}: Hit time of PEs follows the LS time distribution, while the nPE is uniformly distributed between 0 and 10 p.e. This case combines narrow time spread with broad PE-multiplicity, leading to significant degree of waveform overlap, representing mixed-event samples with varying light yields.

\end{itemize}

\subsection{Training and Testing Samples}
\label{sec:data} 

Based on the three types of PE distribution described above, we generate four EMC datasets: UT-UPE, LT-0.1PE, LT-1PE, and LT-UPE. Each dataset comprises 1,000,000 training waveforms, along with 100,000 waveforms each for validation and testing. Unless otherwise specified, both the simulation and reconstruction networks are trained using the same statistics. 

In this study, supervised learning refers to model training that uses the Monte Carlo (MC) truth from the EMC samples. In contrast, weakly supervised learning denotes a training paradigm that does not rely on MC truth; instead, it uses only raw waveforms and coarse initial estimates of PE information.

\section{Structure of the Neural Network}
This chapter describes the network architectures of the Denoising Diffusion Probabilistic Model (DDPM) and the ResNet50 model. The DDPM architecture is based on the U-Net proposed in Ref.~\cite{UNet}, whose theoretical foundation stems from the classical forward diffusion and reverse denoising processes. Building upon this model, we propose an innovative bidirectional conditional diffusion network framework that dynamically exchanges network inputs and conditioning information across different training stages. This design enables the models to synergistically simulate and reconstruct waveforms within a weakly supervised learning paradigm, with iterative refinement progressively enhancing both capabilities. The ResNet50 model adopts the standard architecture and serves primarily as a performance benchmark for nPE reconstruction. Detailed discussions of the training strategies and performance analyses for all networks are presented in Chapters 4 and 5.

\subsection{DDPM}
DDPM is a generative model based on the Markov assumption, comprising a fixed forward noising process $q$ and a reverse denoising process $p_\theta$, with $\theta$ representing the learnable network parameters. Given original data $\mathbf{x_0}$, the forward process gradually corrupts the data into pure Gaussian noise $\mathbf{x_T} \sim \mathcal{N}(0, \mathbf{I})$ over $T$ steps using a predefined variance schedule $\{\beta_t\}_{t=1}^T$. This process is defined by the following Gaussian transition:
\begin{equation}
\label{eq:GausTransfer1}
q(\mathbf{x_t} \mid \mathbf{x_{t-1}}) = \mathcal{N}\bigl(\mathbf{x_t}; \sqrt{1 - \beta_t}\, \mathbf{x_{t-1}},\, \beta_t \mathbf{I} \bigr), \quad t = 1, 2, \dots, T.
\end{equation}
Thanks to the reparameterization trick, noisy samples at any time step $t$ can be directly sampled from $\mathbf{x_0}$:
\begin{equation}
\mathbf{x_t} = \sqrt{\bar{\alpha}_t}\, \mathbf{x_0} + \sqrt{1 - \bar{\alpha}_t}\, \boldsymbol{\epsilon},
\end{equation}
where $\bar{\alpha}_t = \prod_{i=1}^t (1 - \beta_i)$ and $\boldsymbol{\epsilon} \sim \mathcal{N}(0, \mathbf{I})$. The reverse process learns the inverse transformation of the forward process via a neural network, with each step also modeled as a Gaussian distribution:
\begin{equation}
\label{eq:GausTransfer2}
p_{\theta}(\mathbf{x_{t-1}} \mid \mathbf{x_t}) = \mathcal{N}\bigl(\mathbf{x_{t-1}};\, \boldsymbol{\mu}_{\theta}(\mathbf{x_t}, t),\, \boldsymbol{\Sigma}_{\theta}(\mathbf{x_t}, t) \bigr).
\end{equation}
Here, $\boldsymbol{\mu}_{\theta}$ and $\boldsymbol{\Sigma}_{\theta}$ denote the mean and covariance matrix of the conditional distribution $p_\theta(\mathbf{x_{t-1}} \mid \mathbf{x_t})$, respectively. By optimizing the evidence lower bound (ELBO), the DDPM training objective simplifies from predicting the full reverse-process distribution to having the network $\boldsymbol{\epsilon}_{\theta}$ directly predict the noise $\boldsymbol{\epsilon}$ added during the forward process. The simplified loss function is:
\begin{equation}
\label{eq:Loss}
L(\theta) = \mathbb{E}_{\mathbf{x_0}, t, \boldsymbol{\epsilon}} \left[ \bigl\| \boldsymbol{\epsilon} - \boldsymbol{\epsilon}_\theta\bigl( \sqrt{\bar{\alpha}_t}\, \mathbf{x_0} + \sqrt{1 - \bar{\alpha}_t}\, \boldsymbol{\epsilon},\, t \bigr) \bigr\|^2 \right].
\end{equation}
This objective enables the model to synthesize high-quality samples $\mathbf{x_0}$ through iterative denoising starting from pure noise $\mathbf{x_T}$. %The forward noising and reverse denoising processes of DDPM are illustrated in Fig.~\ref{fig:DDPM}.

%\begin{figure*}[t]
%    \centering
%    \includegraphics[width=0.8\linewidth]{DDPM.png}
%    \caption{
%    Schematic illustration of the forward diffusion and reverse denoising processes in DDPM applied to waveform data. 
%    (a) The forward process (left to right) progressively adds Gaussian noise to the original waveform $\mathbf{x_0}$ via a fixed Markov chain, ultimately yielding isotropic Gaussian noise $\mathbf{x_T}$. Each transition is governed by $q(\mathbf{x_t} \mid \mathbf{x_{t-1}})$, where $\mathbf{x_{t-1}}$ and $\mathbf{x_t}$ denote consecutive latent states. 
%    (b) The reverse process (right to left) iteratively denoises the signal using a trained U-Net network $\boldsymbol{\epsilon}_\theta$ to recover the original waveform $\mathbf{x_0}$. Each step is modeled by the learnable distribution $p_\theta(\mathbf{x_{t-1}} \mid \mathbf{x_t})$.
%    }
%    \label{fig:DDPM}
%\end{figure*}

\subsection{Conditional Diffusion Network for Waveform Simulation}
%In practical implementations of DDPM, the reverse denoising network $\boldsymbol{\epsilon}_\theta$ typically adopts a U-Net architecture. This choice is motivated by the fact that the intermediate states $\mathbf{x_t}$ in the diffusion process share the same spatial dimensionality as the input data $\mathbf{x_0}$, and the encoder-decoder structure of U-Net enables multi-scale feature extraction and reconstruction while preserving spatial resolution. 
Compared to the original model in Ref.~\cite{DDPM}, this work systematically modifies the standard U-Net to accommodate the one-dimensional waveform data and the requirements of conditional generation. The detailed network architecture is provided in Table~\ref{tab:Conditional_U-Net}.

First, the network is extended to a conditional U-Net. In addition to the base inputs (noisy waveform $\mathbf{x}$ and diffusion time step $t$), a condition vector $\mathbf{y} = (y_1, y_2, \dots, y_n)$ is introduced, where $y_i$ denotes the number of PEs at time bin $i$. This vector denotes the PE sequence. During training and generation, the condition $\mathbf{y}$ is incorporated at every diffusion step: since DDPM employs a random time-step sampling strategy (i.e., each batch samples a single time step $t$), the network predicts the added noise $\boldsymbol{\epsilon}$ based on $(\mathbf{x}_t, t, \mathbf{y})$ and is optimized by minimizing the mean squared error between predicted and ground-truth noise. To effectively distinguish the data input $\mathbf{x}$ from the condition $\mathbf{y}$, the latter bypasses the initial convolutional layer and is instead mapped through a fully connected (Dense) layer into a 32-dimensional feature vector. This vector is then concatenated with the output of the initial convolution along the channel dimension to achieve feature fusion. Meanwhile, the time step $t$ is transformed into a 256-dimensional vector via sinusoidal positional encoding and processed by a two-layer perceptron to generate a time embedding (\textit{temb}), which serves as a conditional input to every residual block (ResBlock), enabling the network to perceive the current denoising stage.

\begin{figure*}
    \centering
    \includegraphics[width=1.0\linewidth]{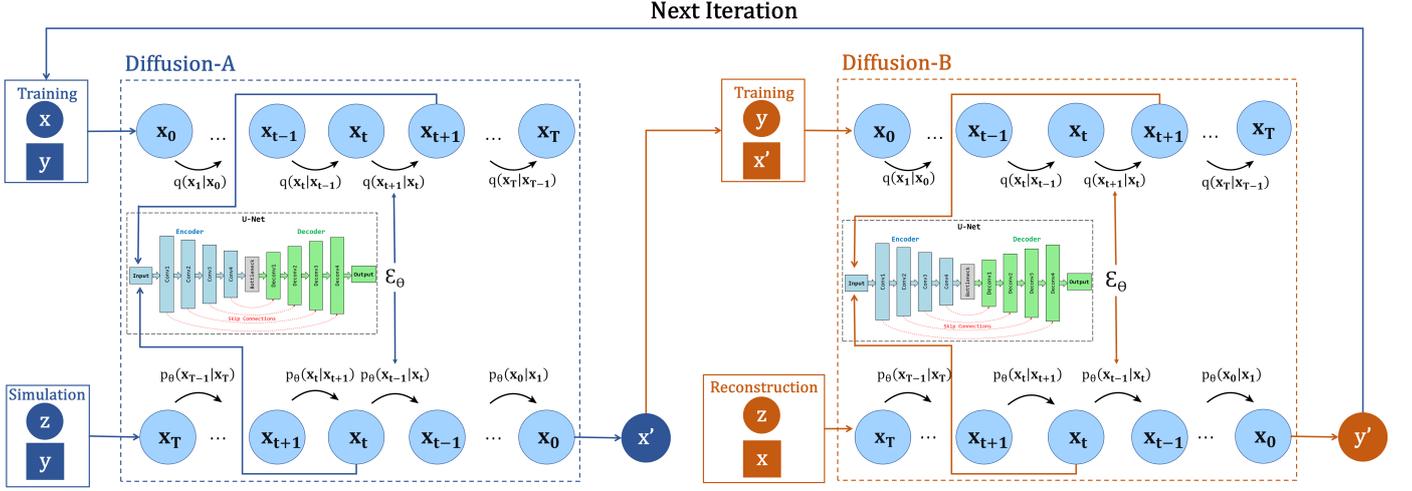}
    \caption{Schematic diagram of the proposed bidirectional conditional diffusion network framework. The framework network consists of two conditional DDPMs: Diffusion-A (left) performs waveform simulation by conditioning the latent vector $\mathbf{z}$ on the PE sequence $\mathbf{y}$, while Diffusion-B (right) reconstructs the PE sequence by conditioning $\mathbf{z}$ on the observed waveform $\mathbf{x}$. Synthetic waveforms $\mathbf{x}'$ generated by DFA are employed to train DFB, and DFB's output $\mathbf{y}'$ is subsequently fed back to retrain DFA, progressively enhancing waveform simulation quality and reconstruction accuracy. The top arrow indicates the iterative training loop between the two diffusion models, enabling self-optimization and joint convergence. }
    \label{fig:Diffusion-A-B}
\end{figure*}

Second, all two-dimensional convolutions are replaced with one-dimensional convolutions. The encoder-decoder structure features four levels of resolution down sampling (1000 $\rightarrow$ 500 $\rightarrow$ 250 $\rightarrow$ 125). Each level comprises a ResBlock and a sampling layer, with an optional attention module (AttentionBlock). Key components are further optimized for the task: The ResBlock is enhanced based on Wide ResNet~\cite{WResNet} and integrates the \textit{temb} to condition its internal computations. Group Normalization (GroupNorm) replaces Batch Normalization to improve stability under small-batch or distributed training; The Swish activation function is uniformly adopted in place of ReLU for better gradient flow; Skip connections are preserved to enable the decoder to effectively utilize high-frequency details extracted by the encoder; Attention-Blocks are deployed only at higher-level resolutions (250 and 125) to model long-range time dependencies in the waveform sequence; Down sampling uses strided Conv1D with stride 2, while up-sampling is implemented via nearest-neighbor interpolation followed by convolution to avoid checkerboard artifacts commonly associated with transposed convolutions.

In summary, the key innovation of this U-Net architecture lies in its multi-source conditioning design, which allows the network to simultaneously perceive the current denoising state ($\mathbf{x}$), diffusion progress ($t$), and target physical properties ($\mathbf{y}$). This design forms the foundation of the bidirectional conditional diffusion framework proposed in this work.

\subsection{Bidirectional Conditional Diffusion Network for Waveform Simulation and Reconstruction}
\label{sec:BidiCondiDiffNetwork}

The Bidirectional Conditional Denoising Diffusion Probabilistic Model (BCDDPM) consists of two structurally identical conditional DDPMs. The PE-conditioned diffusion network (Diffusion-A, DFA) synthesizes waveforms conditioned on a PE sequence $\mathbf{y}$, while the waveform-conditioned diffusion network (Diffusion-B, DFB) synthesizes PE sequences conditioned on a waveform $\mathbf{x}$. This dual-module architecture enables synergistic waveform simulation and reconstruction. The overall framework is illustrated in Fig.~\ref{fig:Diffusion-A-B}.

Specifically, DFA is trained to synthesize waveforms $\mathbf{x}$ using the PE sequence $\mathbf{y}$ as a conditioning input, with the diffusion time step $t$ (indicating the current denoising stage) as an auxiliary input. Since $\mathbf{y}$ is injected at every diffusion step, the reverse denoising process of DFA is expressed as:
\begin{equation}
p_\theta(\mathbf{x_{0:T}} \mid \mathbf{y}) = p_\theta(\mathbf{x_T}) \prod_{t=1}^T p_\theta(\mathbf{x_{t-1}} \mid \mathbf{x_t}, \mathbf{y}).
\end{equation}
After training, DFA can synthesize waveforms $\mathbf{x}' \sim p_\theta(\mathbf{x} \mid \mathbf{y})$ given a latent vector $\mathbf{z}$ and a PE condition $\mathbf{y}$.

Conversely, DFB is trained to reconstruct PE sequences $\mathbf{y}$ using waveforms synthesized by DFA ($\mathbf{x}'$) as the conditioning input. In the network implementation, $\mathbf{y}$ is processed directly through the initial convolutional layer, whereas $\mathbf{x}'$ is first mapped via a Dense layer before being fused with the convolutional features. Similarly, the waveform condition $\mathbf{x}'$ is incorporated at each diffusion step, yielding the reverse process:
\begin{equation}
p_\theta(\mathbf{y_{0:T}} \mid \mathbf{x}') = p_\theta(\mathbf{y_T}) \prod_{t=1}^T p_\theta(\mathbf{y_{t-1}} \mid \mathbf{y_t}, \mathbf{x}').
\end{equation}
Upon convergence, DFB can reconstruct PE sequences $\mathbf{y}' \sim p_\theta(\mathbf{y} \mid \mathbf{x})$ from observed waveforms $\mathbf{x}$ and a redundant latent vector $\mathbf{z}$.

In the weakly supervised learning paradigm (where ground-truth PE sequences are unavailable), DFB may produce a refined estimate of PE sequences $\mathbf{y}'$ that is more accurate than the initial condition $\mathbf{y}$. This improved $\mathbf{y}'$ can then be used to retrain DFA, enhancing waveform simulation quality. This iterative refinement continues until both waveform quality and reconstruction accuracy converge simultaneously. Consequently, BCDDPM adopts this iterative training strategy under weak supervision, enabling continuous self-improvement. The detailed procedure is described in Section~\ref{sec:looptrain}.

\subsection{ResNet50}
ResNet50 is a widely used deep convolutional neural network comprising 50 layers. Its core innovation lies in the residual learning framework, which mitigates the vanishing gradient problem in deep networks through skip connections. Each residual block consists of multiple $3\times3$ convolutional layers and batch normalization, enabling the network to learn rich hierarchical feature representations. In this work, ResNet50 serves as a baseline model for evaluating the waveform simulation and reconstruction performance of BCDDPM. Given its limited capability in reconstructing full PE sequences, we restrict its application to the nPE classification task.

To adapt ResNet50 to one-dimensional waveform data, all two-dimensional convolutions and pooling operations in the original architecture~\cite{ResNet50} are replaced with their one-dimensional counterparts. Specifically, the structure and depth of the bottleneck residual blocks remain unchanged, with only the spatial dimension of convolutional kernels reduced from 2D to 1D (e.g., $3\times3$ kernels are replaced by $3\times1$ kernels). Additionally, the output layer employs global average pooling followed by a fully connected layer to produce multi-class outputs aligned with the label dimension.

\section{Supervised Learning Approach}
Previous study~\cite{MLPhoton} have demonstrated that supervised training of RawNet effectively predicts nPE and improves energy resolution of JUNO by approximately 2-3\%. Supervised learning serves as both an idealized performance ceiling and a rigorous benchmark for assessing a network's task suitability. This section investigates the waveform simulation and reconstruction performance of diffusion models under supervised learning, benchmarked against ResNet50 under identical training conditions. Their performance under weakly supervised learning will be examined in the following chapter.

\subsection{Training Methods}
\label{sec:sup-train}

The hyperparameter configuration for BCDDPM is summarized in Table~\ref{tab:DDPM_hyperparams}. All parameters are selected and validated based on experiments. In this work, the number of diffusion steps is set to $T = 200$, with different noise scheduling strategies employed for DFA and DFB to achieve high-quality waveform simulation and reconstruction within a limited number of steps. Specifically, DFA adopts a broad noise schedule range $[10^{-4}, 10^{-1}]$, which facilitates the progressive recovery of both the overall waveform envelope and fine temporal structures during reverse denoising; DFB utilizes a narrower schedule interval $[10^{-3}, 10^{-2}]$, ensuring stable denoising dynamics while focusing on precise identification of PE features from noisy sequences.
To address potential numerical instabilities arising from unnormalized training data, such as gradient explosion or unphysical oscillations, a bounded clipping constraint $[-10^{2}, 10^{3}]$ is applied to the network outputs. This constraint significantly enhances training stability. All training-related hyperparameters are optimized via a coarse-to-fine grid search under supervised learning. The loss function follows  Eq.~\ref{eq:Loss}, which minimizes the mean squared error between the predicted and ground-truth noise.

The hyperparameter configuration for ResNet50 is detailed in Table~\ref{tab:ResNet50_hyperparams}. The network is trained using the AdamW optimizer with a weight decay coefficient of $10^{-4}$, which balances convergence stability and regularization effectiveness in deep residual architecture. The initial learning rate is set to $1 \times 10^{-3}$, consistent with the typical range recommended for AdamW. A dynamic learning rate decay schedule is further employed to enable finer convergence during the later stages of training. Additionally, lightweight gradient clipping is applied to suppress anomalous gradient spikes, and a large batch size is adopted. This design significantly enhances numerical stability in distributed training environments and improves the reliability of batch normalization statistics. The loss function is implemented as sparse categorical crossentropy.

In the following, the DFA models trained on the four EMC datasets described in Section~\ref{sec:data} and their corresponding MC truth labels are denoted as: \textbf{UT-UPE-DFA}, \textbf{LT-UPE-DFA}, \textbf{LT-1PE-DFA}, and \textbf{LT-0.1PE-DFA}. 

The DFB models are trained on LT-UPE samples generated either from the EMC or from the above four DFA models, and are labeled as: \textbf{LT-UPE-EMC-DFB}, \textbf{UT-UPE-DFA-DFB}, \textbf{LT-UPE-DFA-DFB}, \textbf{LT-1PE-DFA-DFB}, and \textbf{LT-0.1PE-DFA-DFB}. 

Similarly, ResNet-50 models trained on LT-UPE samples from the EMC and from the LT-UPE-DFA model are denoted as: \textbf{LT-UPE-EMC-RN} and \textbf{LT-UPE-DFA-RN}.

\subsubsection{Diffusion-A}
\label{sec:Diffusion-A}

Fig.~\ref{fig:AllLossCurves} shows the evolution of training and validation losses during supervised learning for all DFA models.  The final model weights used in subsequent analyses are selected based on the minimum validation loss. After excluding the first 10 epochs to mitigate early-training instability, the epochs corresponding to the lowest validation loss for each model are marked by stars in Fig.~\ref{fig:AllLossCurves}. This selection strategy is justified by two considerations: (1) the minimum validation loss indicates an optimal balance between model complexity and generalization, thereby mitigating overfitting; (2) all selected epochs lie within the stable convergence regime, where both training and validation losses plateau, indicating near-optimal data fitting.

\subsubsection{Diffusion-B}
The primary task of DFB is nPE reconstruction, which requires training data with uniformly distributed nPE. To this end, we train DFB on LT-UPE samples generated either from the EMC or from the four DFA models described in Section~\ref{sec:Diffusion-A}, using identical PE sequences as in the EMC.

Fig.~\ref{fig:AllLossCurves} compares the training and validation loss evolution across different DFB models. For subsequent analysis, we select the model weights with the lowest validation loss; the corresponding epochs are indicated by stars in Fig.~\ref{fig:AllLossCurves}. This approach avoids early-training instability and captures stable convergence behavior. The validation losses are ordered as 
LT-UPE-EMC-DFB < UT-UPE-DFA-DFB < LT-UPE-DFA-DFB < LT-1PE-DFA-DFB < LT-0.1PE-DFA-DFB.

\subsubsection{ResNet50}
\begin{figure}[t]
    \centering
    \includegraphics[width=0.8\linewidth]{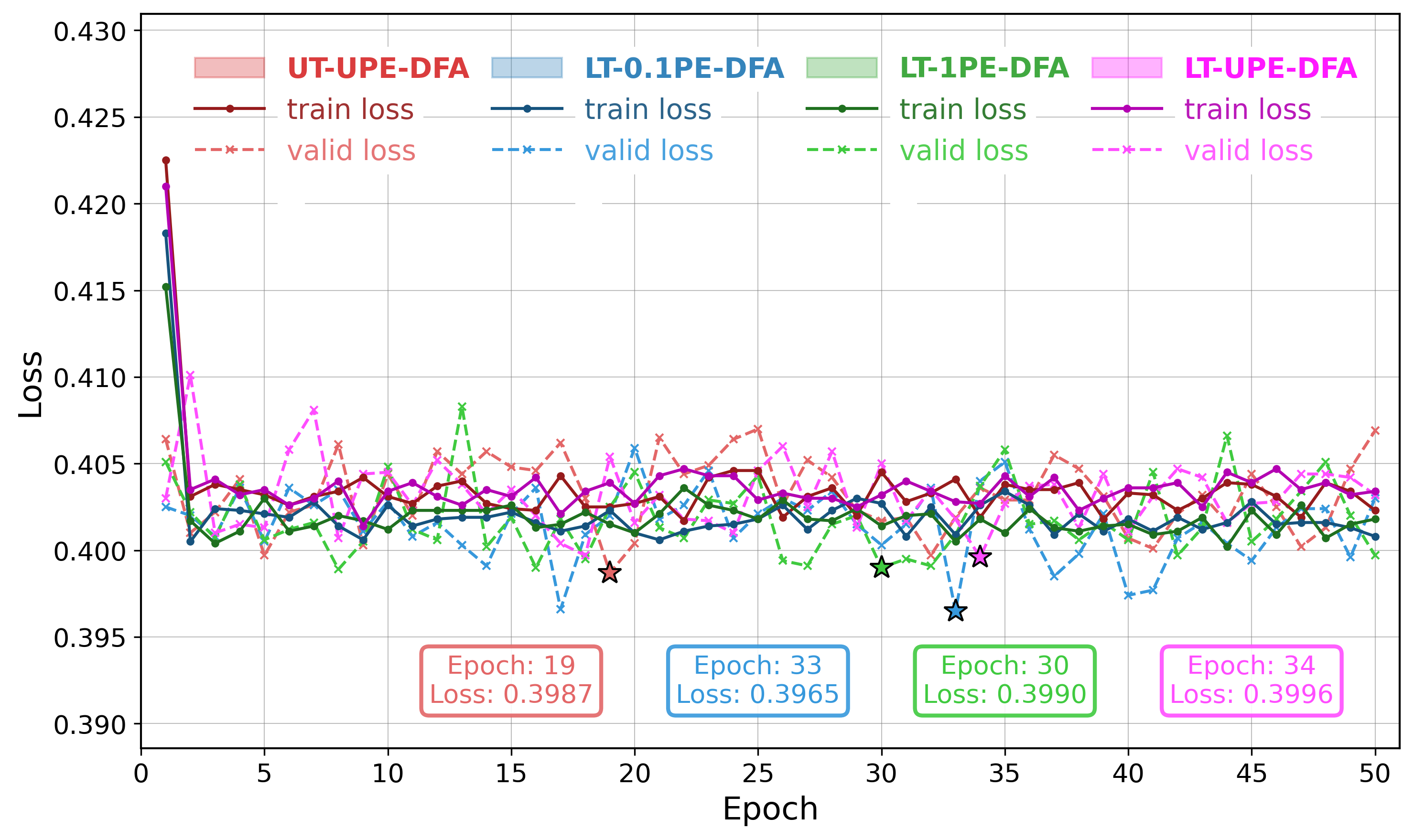}
    \includegraphics[width=0.8\linewidth]{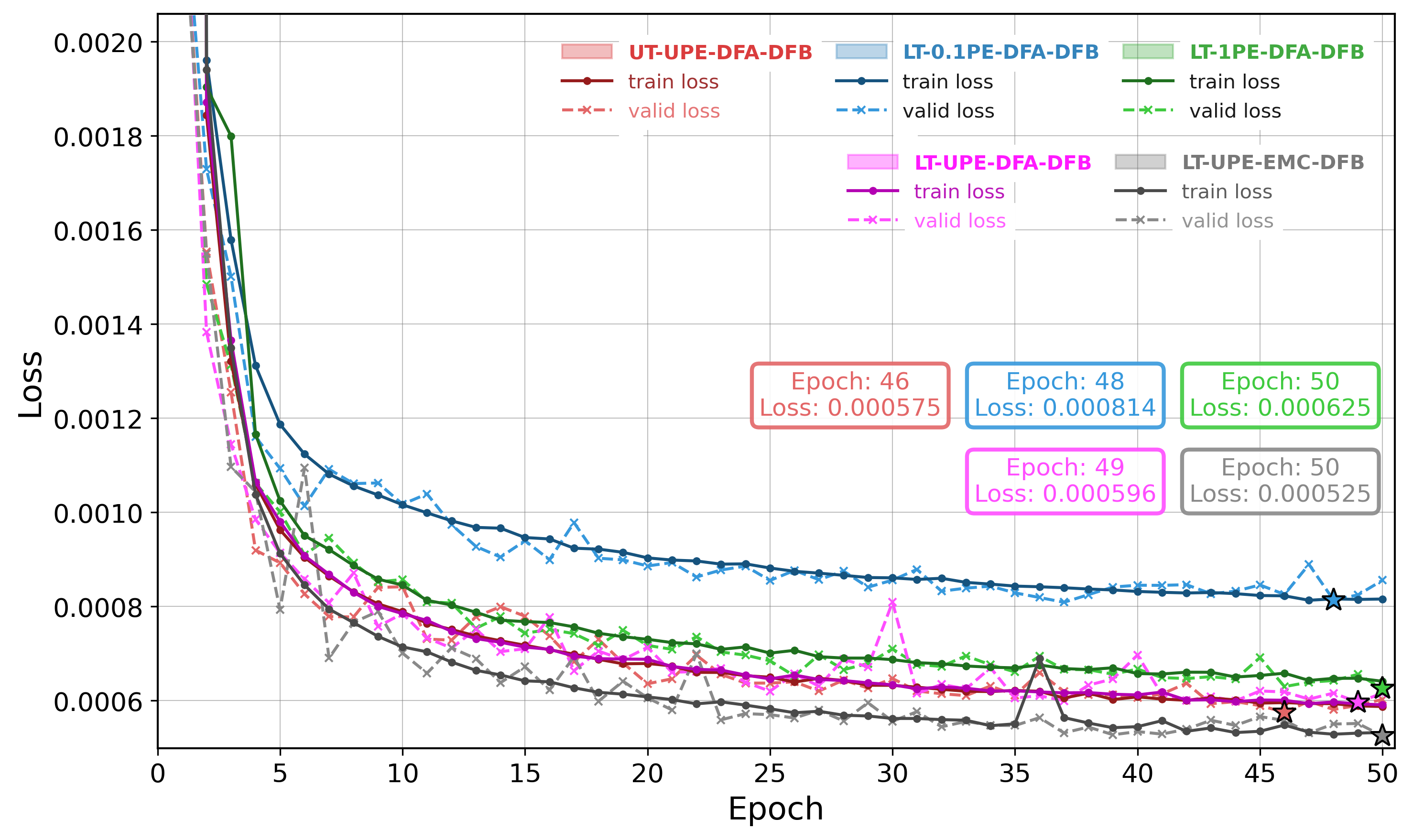}
    \includegraphics[width=0.8\linewidth]{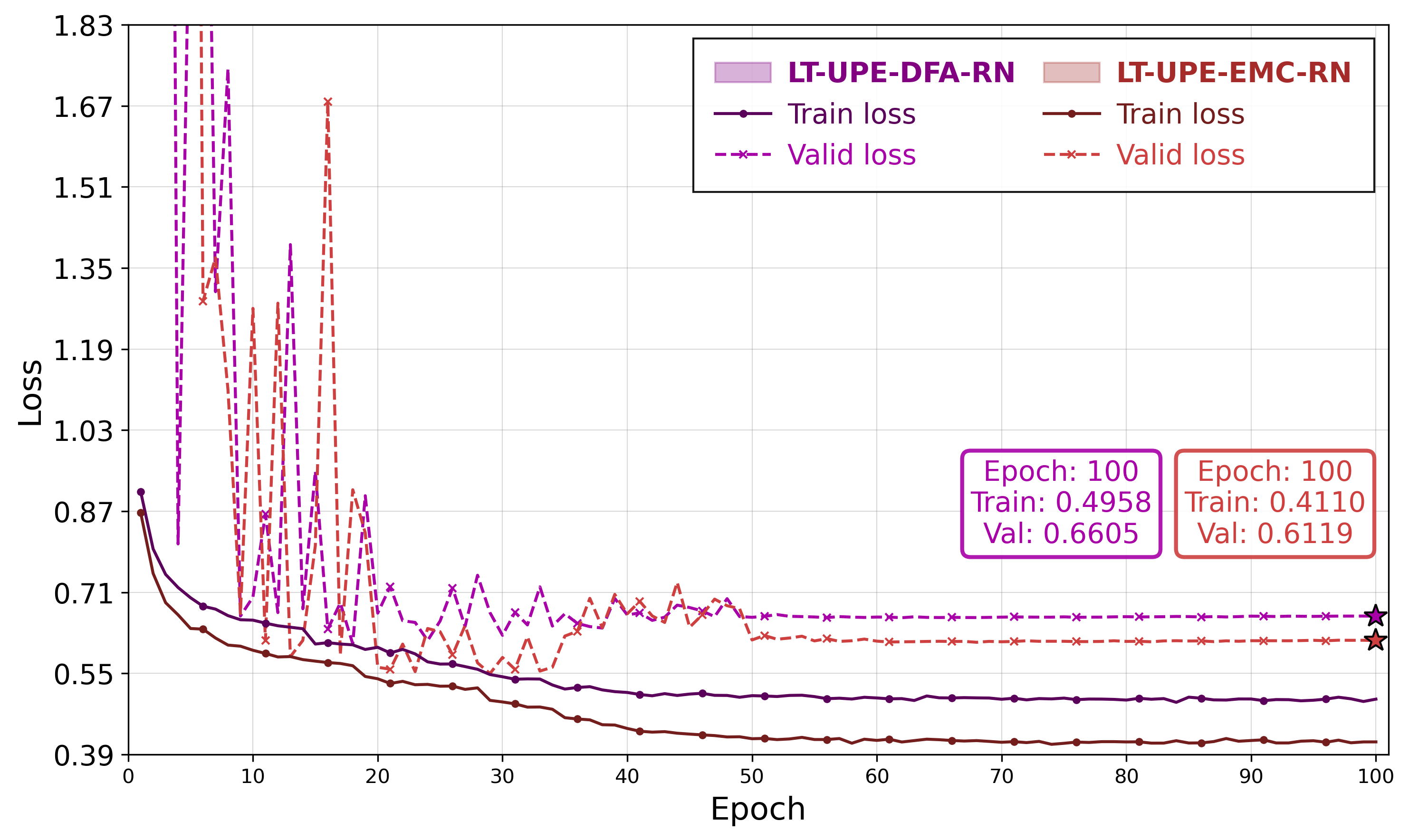}
    \caption{Comparison of training and validation loss curves for different models: DFA(top), DFB(middle), and ResNet50(bottom). All curves highlight key performance points marked by asterisks.}
    \label{fig:AllLossCurves}
\end{figure}

ResNet50 is trained on LT-UPE samples generated by both the EMC and the LT-UPE-DFA model. The evolution of its training and validation losses is shown in Fig.~\ref{fig:AllLossCurves}. Since both losses stabilize after epoch 50 without exhibiting the rising validation loss indicative of overfitting, we select the ResNet50 model at epoch 100 for subsequent analysis.

\subsection{Performance Analysis}
\subsubsection{Waveform Simulation}
The performance of the simulation network is primarily evaluated by comparing the shapes of synthetic waveforms with those of real waveforms, with a focus on sPE waveforms and overlapping waveforms. This comparison is conducted from two aspects: averaged waveform and charge spectrum. In particular, the charge spectrum is quantified using two key characteristics: its mean and standard deviation.

The charge spectrum, defined as the distribution of waveform areas, serves as a critical metric for assessing waveform simulation quality. To mitigate the impact of white noise on charge estimation, we first apply the peak-finding algorithm described in Section~\ref{sec:xunfeng} to determine whether a waveform contains PEs. For waveforms with identified signals, integration windows are defined based on peak positions. Specifically, for the $i$-th detected peak at time $t^{\mathrm{p}}_{i}$, the integration window is set to $[t^{\mathrm{p}}_{i} - 30,\, t^{\mathrm{p}}_{i} + 50]~\mathrm{ns}$. Overlapping integration intervals are merged to avoid double-counting. The charge is then obtained by integrating the waveform over the selected interval(s). If no signal is detected, the charge is assigned zero.

\begin{figure*}
    \centering
    \begin{subfigure}[b]{0.33\linewidth}  
        \centering
        \includegraphics[width=\linewidth,  height=10cm]{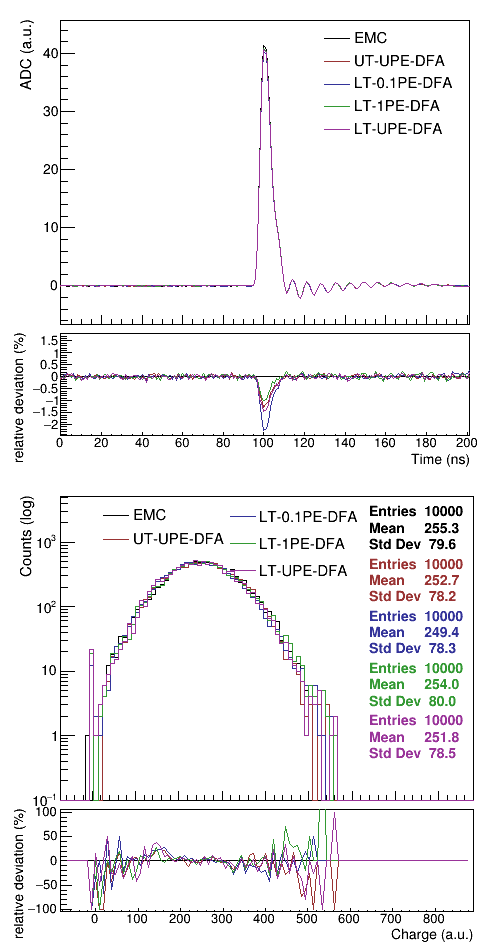}
        \caption{sPE samples}
        \label{fig:DiffA_Check_a}
    \end{subfigure}
    \hfill  
    \begin{subfigure}[b]{0.33\linewidth}
        \centering
        \includegraphics[width=\linewidth, height=10cm]{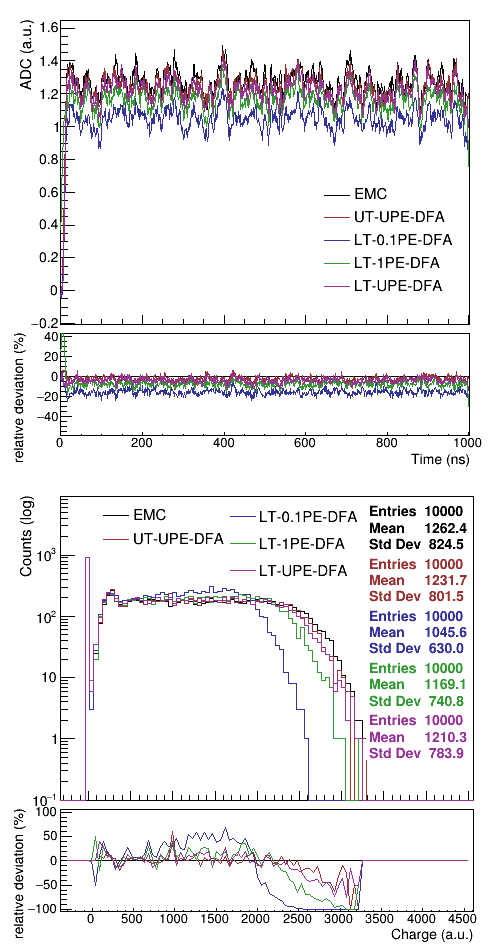}
        \caption{UT-UPE samples}
        \label{fig:DiffA_Check_b}
    \end{subfigure}
    \hfill
    \begin{subfigure}[b]{0.33\linewidth}
        \centering
        \includegraphics[width=\linewidth, height=10cm]{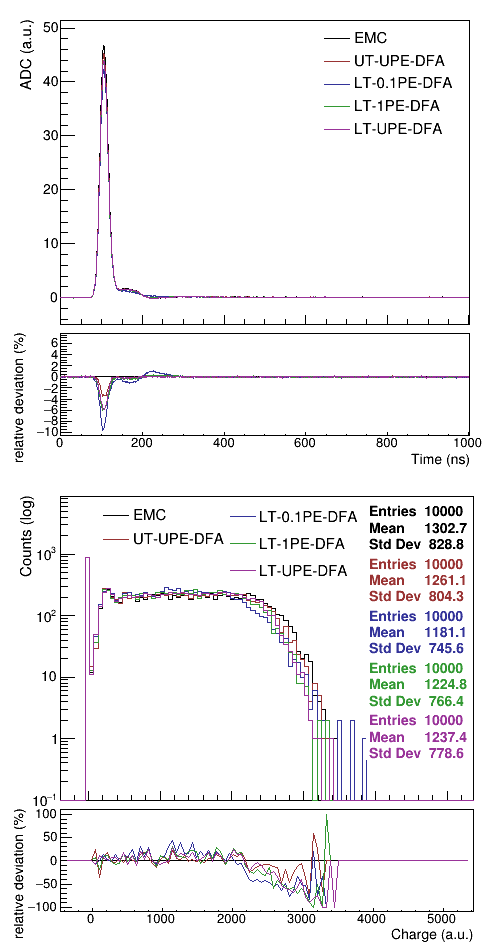}
        \caption{LT-UPE samples}
        \label{fig:DiffA_Check_c}
    \end{subfigure}
    \caption{
        Comparison of averaged waveforms and charge spectra between EMC and DFA models. The relative deviation deviations are defined relative to EMC.
        (a) sPE samples: All four models exhibit waveform deviations below 2\% relative to EMC, with both the mean and resolution deviations of the charge spectra under 2\%, demonstrating accurate reproduction of characteristics of sPE charge. 
        (b) UT-UPE samples: Synthetic waveforms show uniform temporal distribution across all models, with overall waveform shapes consistent with EMC, indicating successful learning of the strong correlation between PE times and peak positions. Comparisons of charge spectra reveals that LT-0.1PE-DFA's waveform deviation primarily stems from systematically lower charges in the high nPE region. 
        (c) LT-UPE samples: Compared to EMC, UT-UPE-DFA, LT-UPE-DFA, and LT-1PE-DFA show waveform deviations below 6\%, while LT-0.1PE-DFA exhibits a maximum deviation of approximately 10\%. Relative to the UT-UPE case in (b), all models demonstrate reduced deviation in both waveform and charge spectrum for LT-UPE samples.
    }
    \label{fig:DiffA_Check}
\end{figure*}

The averaged sPE waveforms and charge spectra from the EMC and DFA models are shown in Fig.~\ref{fig:DiffA_Check_a}. All four models successfully learn the characteristic features of sPE waveform, with deviations below 2\% relative to EMC. Both the mean and resolution deviations of the charge spectrum are also less than 2\%, demonstrating that the models accurately reproduce the statistical properties of the sPE charge. This confirms the method's capability to generate precise and smooth sPE charge spectrum.

Fig.~\ref{fig:DiffA_Check_b} compares the averaged waveforms and charge spectra for UT-UPE samples from the EMC and DFA models. The synthetic waveforms exhibit uniform temporal distribution across all four models, indicating successful learning of the strong correlation between PE times and waveform peak positions. The waveform deviations are 2.5\% (UT-UPE-DFA), 18.2\% (LT-0.1PE-DFA), 8.2\% (LT-1PE-DFA), and 4.4\% (LT-UPE-DFA), respectively. The charge spectrum comparison in Fig.~\ref{fig:DiffA_Check_b} reveals that LT-0.1PE-DFA tends to synthesize waveforms with smaller charges in the high nPE region, which is the primary cause of its significantly lower averaged waveform.

The averaged waveforms and charge spectra for LT-UPE samples from the EMC and DFA models are presented in Fig.~\ref{fig:DiffA_Check_c}. The deviations of peak amplitude are 3.3\% (UT-UPE-DFA), 9.5\% (LT-0.1PE-DFA), 5.8\% (LT-1PE-DFA), and 5.9\% (LT-UPE-DFA), indicating that UT-UPE-DFA, LT-1PE-DFA, and LT-UPE-DFA effectively capture the relationship between dense multi-PE and waveform features. Comparing Figs.~\ref{fig:DiffA_Check_b} and~\ref{fig:DiffA_Check_c}, we observe that LT-0.1PE-DFA synthesizes higher-quality LT-UPE waveforms than UT-UPE waveforms, while the other three models show comparable performance between the two sample types. This suggests that LT-0.1PE-DFA has learned features of overlapping waveforms specific to the LS time distribution, but these features do not generalize to other temporal configurations.

\begin{figure*}
    \centering
    \includegraphics[width=0.95\linewidth]{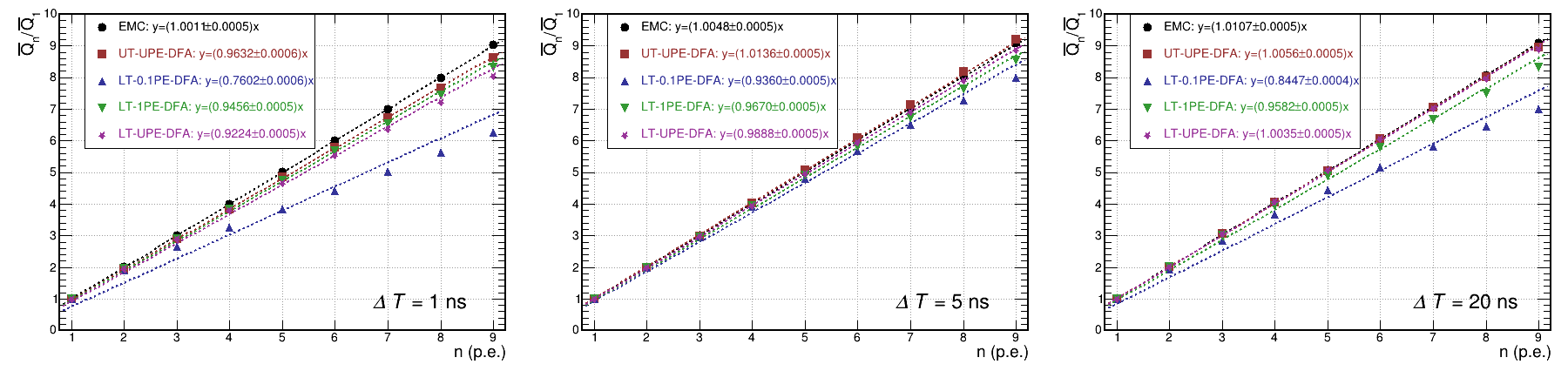}
    \includegraphics[width=0.95\linewidth]{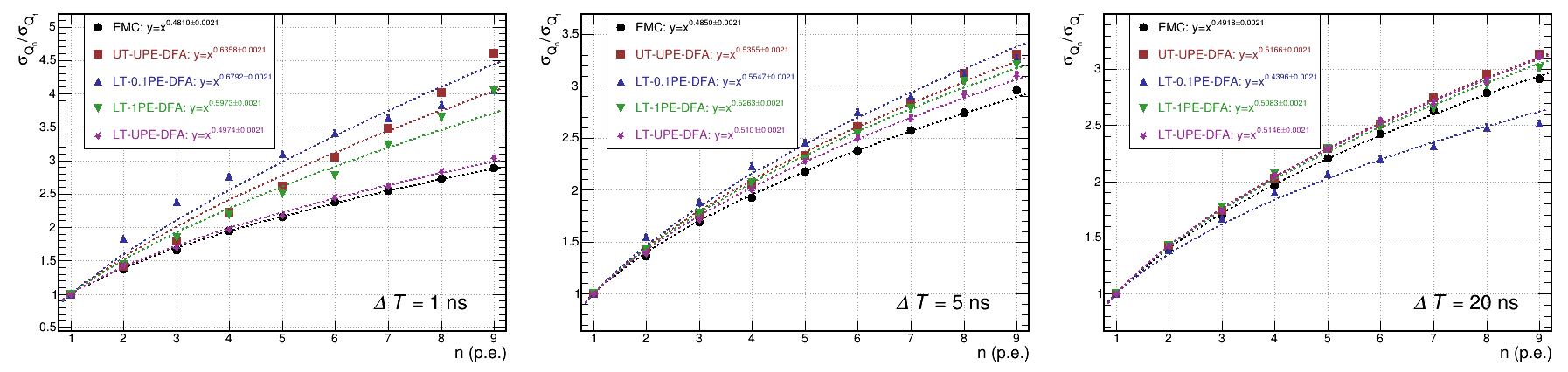}
    \caption{
        The upper panel shows the charge linearity comparison between EMC and DFA models at different $\Delta T$: $\bar{Q}_n/\bar{Q}_1$ as a function of nPE $n$. Dashed lines show linear fits. The lower panel shows the charge resolution comparison between EMC and DFA models at different $\Delta T$: $\sigma_{Q_n}/\sigma_{Q_1}$ as a function of $n$. Dashed lines indicate power-law fits.
    }
    \label{fig:Qlinearity-Res}
\end{figure*}

Under an ideal Poisson process, the mean charge $\bar{Q}_n$ of waveforms containing $n$ PEs follows a linear relationship $\bar{Q}_n/\bar{Q}_1 = n$, while the charge resolution $\sigma_{Q_n}$ satisfies $\sigma_{Q_n}/\sigma_{Q_1} = \sqrt{n}$, where $\sigma_{Q_n}$ denotes the standard deviation of the charge spectrum for n PEs. To investigate how charge linearity and resolution of synthetic waveforms depend on nPE, we generated waveform samples with varying nPE using both EMC and DFA, while systematically varying the time interval $\Delta T$ between PEs in the conditioning vector (smaller $\Delta T$ corresponds to more severe waveform overlap). 

Fig.~\ref{fig:Qlinearity-Res} shows the dependence of $\bar{Q}_n$ and $\sigma_{Q_n}$ on $n$ for $\Delta T$ = 1, 5, and 20 ns. The results indicate that all datasets exhibit approximately linear relationship between charge and nPE, and the resolution follows an approximate power-law relationship. However, both quantities show clear dependence on $\Delta T$, deviating from theoretical expectations. 

\begin{figure}[t]
     \centering

    \includegraphics[width=0.95\linewidth]{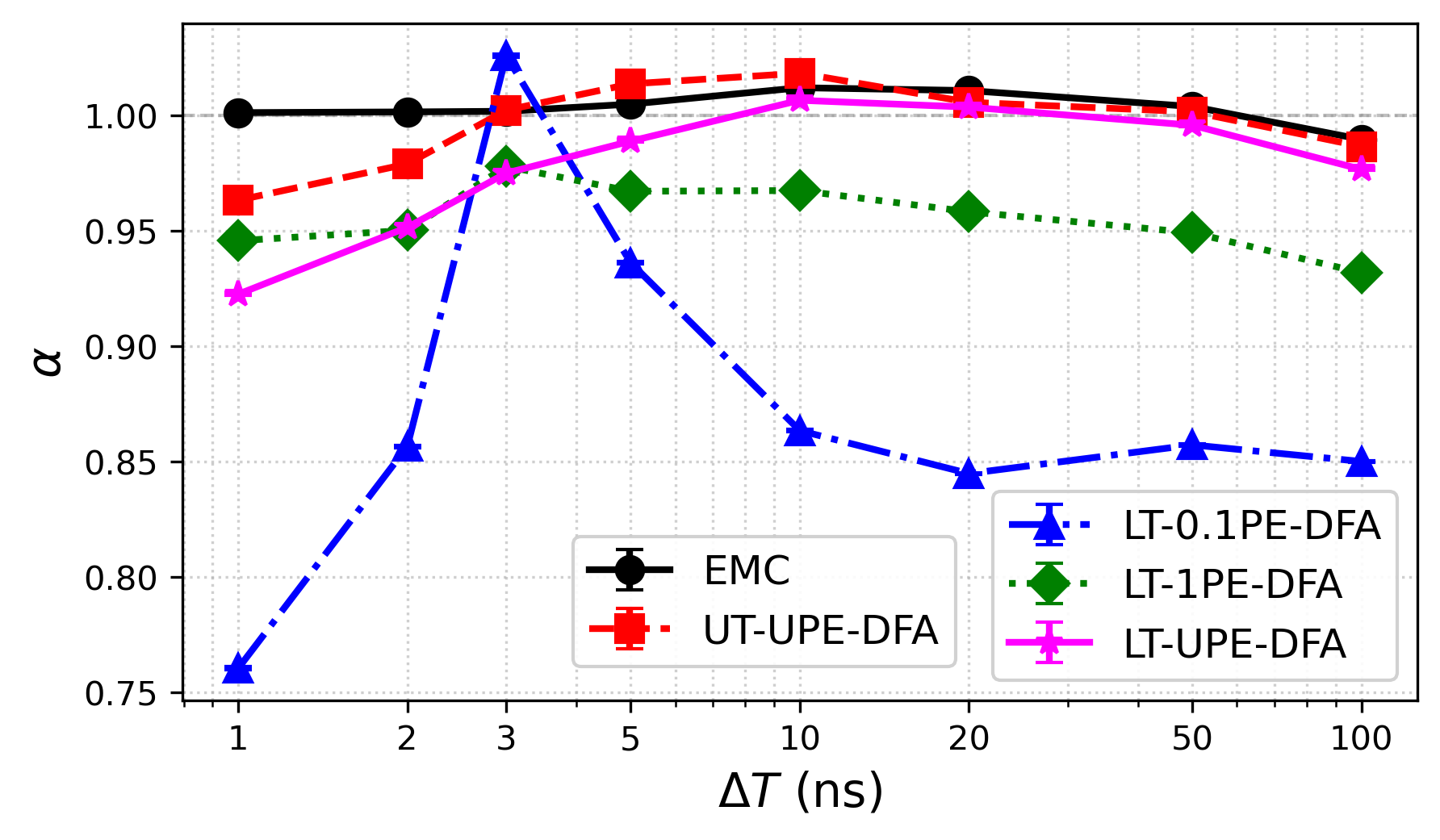}
    \includegraphics[width=0.95\linewidth]{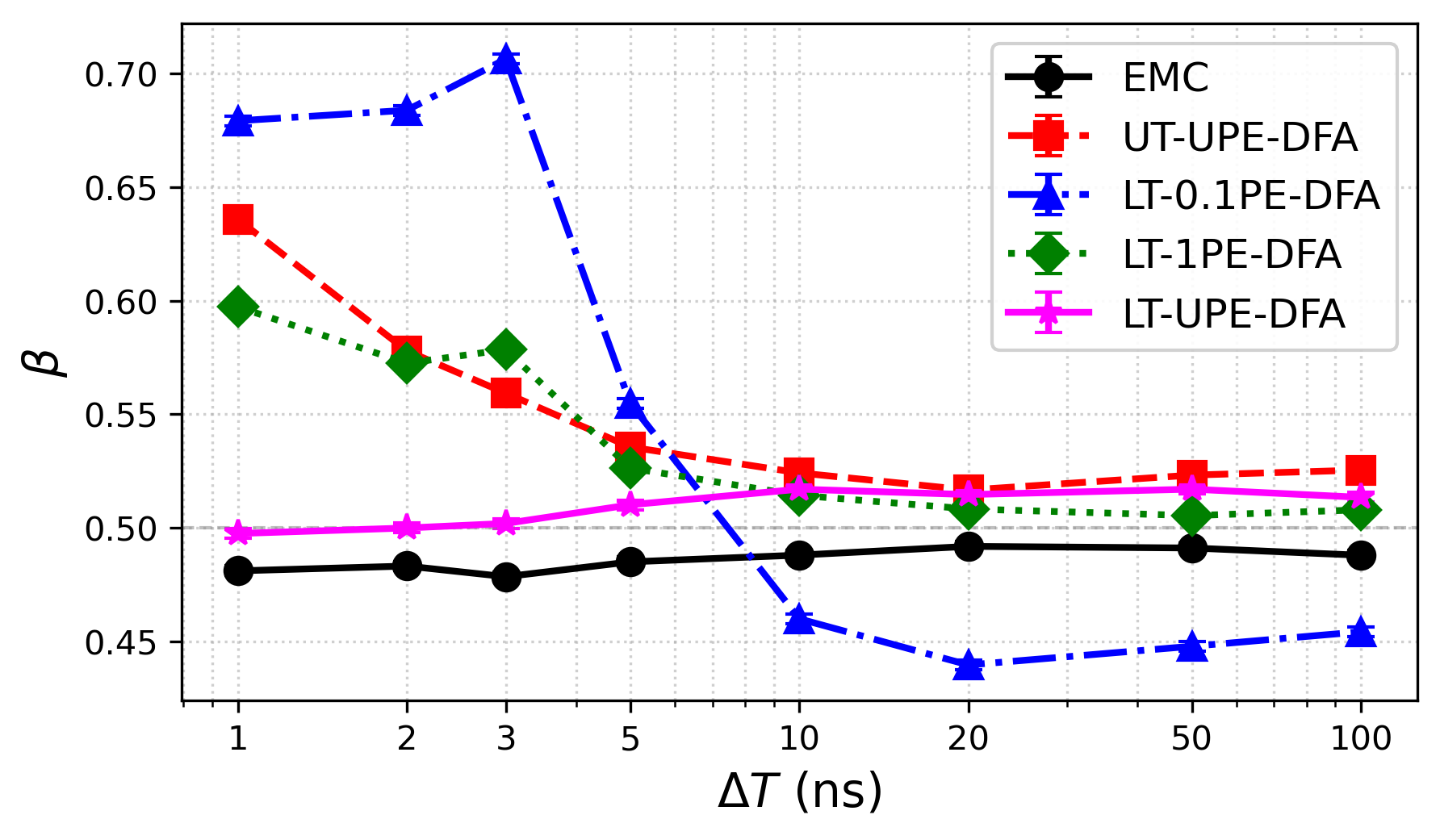}

    \caption{
        Dependence of the charge linearity coefficient $\alpha$ (upper) and resolution exponent $\beta$ (lower) on $\Delta T$ for EMC and DFA models. The EMC represents the ideal case, with $\alpha \approx 1.00$ and $\beta \approx 0.48$, consistent with theoretical expectations. Both LT-UPE-DFA and UT-UPE-DFA maintain charge linearity and resolution consistent with EMC across all $\Delta T$ values. UT-UPE-DFA achieves near-ideal $\alpha$, but exhibits elevated $\beta$ in the severe overlap regime ($\Delta T \leq 5~\mathrm{ns}$). LT-UPE-DFA shows $\beta$ closest to 0.5 for $\Delta T \leq 5~\mathrm{ns}$, indicating more accurate modeling of charge variance. LT-1PE-DFA demonstrates intermediate performance, with $\alpha \in [0.94, 0.97]$ and $\beta$ approaching 0.5 for $\Delta T \geq 10~\mathrm{ns}$. LT-0.1PE-DFA shows significant deviations at both short ($\Delta T \leq 3~\mathrm{ns}$) and long ($\Delta T \geq 10~\mathrm{ns}$) intervals, reflecting its inability to accurately model charge distributions due to the scarcity of multi-PE waveforms in its training data.
}
    \label{fig:QdTfit}
\end{figure}

To quantify model performance, we fit the relationships using $\bar{Q}_n/\bar{Q}_1 = \alpha\,n$ and $\sigma_{Q_n}/\sigma_{Q_1} = n^{\beta}$, defining $\alpha$ (linearity coefficient) and $\beta$ (resolution exponent) as key metrics. The fitted values of $\alpha$ and $\beta$ as functions of $\Delta T$ are shown in Fig.~\ref{fig:QdTfit}. The EMC represents the ideal case, with $\alpha \approx 1.00$ and $\beta \approx 0.48$ (close to the theoretical value of 0.50). The characteristics of the four DFA models are summarized as follows:

\begin{itemize}
    \item \textbf{LT-UPE-DFA and UT-UPE-DFA} maintain charge linearity and resolution consistent with EMC across all $\Delta T$ values. LT-UPE-DFA shows $\alpha \in [0.92, 1.00]$ (slightly low) and $\beta \in [0.50, 0.52]$, indicating accurate learning of characteristics of charge even for severely overlapping waveforms. UT-UPE-DFA achieves $\alpha\in [0.96, 1.01]$ (near-ideal linearity), but exhibits elevated $\beta$ (by $\sim$7\% at 5 ns) for $\Delta T \leq 5$ ns, suggesting insufficient learning of charge variance characteristics under strong overlap conditions.
    
    \item \textbf{LT-0.1PE-DFA} shows significantly reduced $\alpha$ for $\Delta T \leq 2$ ns and $\Delta T \geq 10$ ns (with $n > 3$ p.e.), indicating systematic underestimation of charge in both dense and sparse multi-PE scenarios. Its $\beta$ is notably elevated for $\Delta T \leq 5$ ns, reflecting inadequate learning of charge variance, likely due to the scarcity of multi-PE waveforms in its training data.
    
    \item \textbf{LT-1PE-DFA} exhibits intermediate behavior between LT-0.1PE-DFA and UT-UPE-DFA. It maintains $\alpha \in [0.94, 0.97]$ (slightly below) across all $\Delta T$ values. In terms of charge variance modeling, LT-1PE-DFA slightly outperforms UT-UPE-DFA, with $\beta$ approaching 0.5 more closely for $\Delta T \geq 10$ ns.
\end{itemize}

In summary, LT-UPE-DFA achieves the best overall performance in generating sPE, sparse multi-PE, and dense multi-PE waveforms. UT-UPE-DFA shows weaker capability in modeling severely overlapping waveforms, while LT-0.1PE-DFA performs worst in both sparse and dense multi-PE scenarios. LT-1PE-DFA exhibits intermediate behavior between LT-UPE-DFA and LT-0.1PE-DFA, with overall performance comparable to LT-UPE-DFA. These results indicate that the diffusion network's ability to learn the characteristics of overlapping waveform is positively correlated with the frequency of such characteristics in the training samples. Furthermore, when the target feature (e.g., charge distribution of overlapping waveforms) exhibits weak correlation with the conditioning input (e.g., PE sequence), the model requires more training samples containing that feature to learn effectively. The next subsection analyzes how DFA performance impacts DFB reconstruction quality.

\subsubsection{Waveform Reconstruction}
\begin{figure}
    \centering
    %\begin{subfigure}[b]{0.95\linewidth}
    %    \centering
        \includegraphics[width=0.95\linewidth]{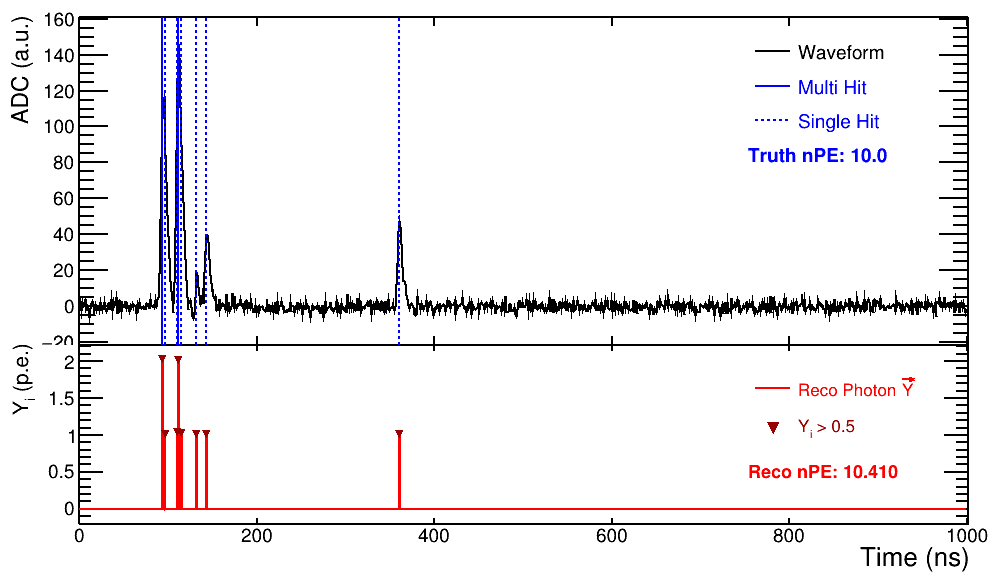}
        \includegraphics[width=0.95\linewidth]{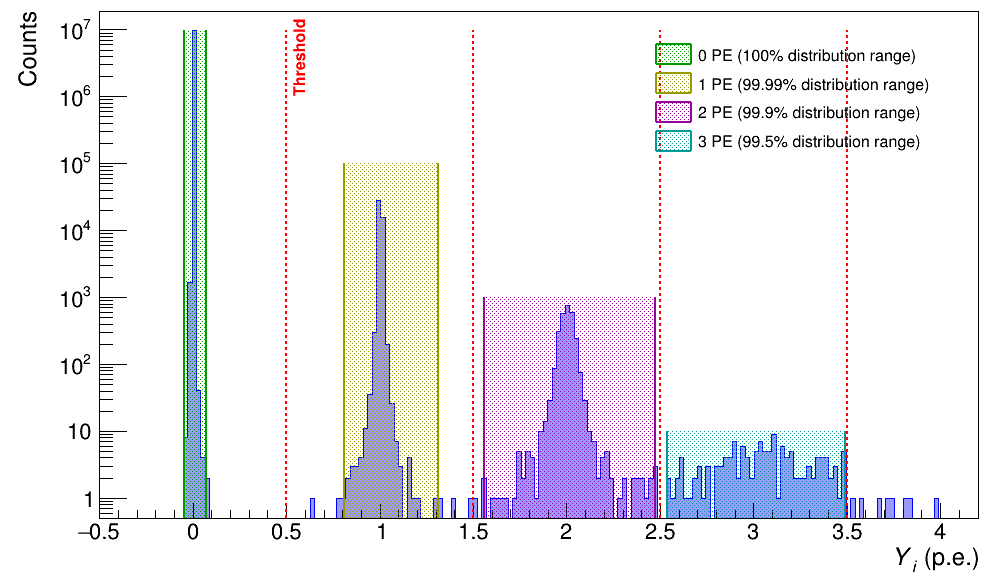}
    %    \caption{Histogram of DFB output values $Y_i$, showing peaks centered at integer values (shaded regions) corresponding to 0, 1, 2, and 3 PEs. Red dashed lines indicate the classification thresholds for rounding.}
    %$$    \label{fig:WaveRec-b}
    %$\end{subfigure}
    
    \caption{Upper panel: A typical result of DFB waveform reconstruction. Lower panel: Histogram of DFB output values $Y_i$, showing peaks centered at integer values (shaded regions) corresponding to 0, 1, 2, and 3 PEs. Red dashed lines indicate the classification thresholds for rounding.
    }
    \label{fig:WaveRec}
\end{figure}

A  typical result of DFB waveform reconstruction is shown in Fig.~\ref{fig:WaveRec}. The DFB output is a reconstructed PE sequence vector $\mathbf{Y} = (Y_1, Y_2, \dots, Y_{1000})$, where $Y_i$ is a floating value representing the estimated nPE at time bin $i$. As shown in Fig.~\ref{fig:WaveRec}, most $Y_i$ are near zero. We therefore adopt $Y_i > 0.5$ as the threshold for identifying PEs at time bin $i$, and define the reconstructed nPE at that bin as $k_i = \lfloor Y_i + 0.5 \rfloor$. The total reconstructed nPE for the waveform is then $k = \sum_{i=1}^{1000} k_i$.

Reconstruction performance is evaluated from two aspects: nPE and timing precision. The former is defined by the conditional probability $P(k|n)$, the probability of reconstructing $k$ PEs given $n$ true PEs. For $n$ PEs events, we define the reconstructed mean as $\bar{k} = \sum_{k=0}^{\infty} k \, P(k\mid  n)$ and the resolution as $\sigma_k = \sqrt{\sum_{k=0}^{\infty} (k - \bar{k})^2 \, P(k\mid n)}$, which quantify the bias ((i.e., $\bar{k}-n$) and resolution of nPE reconstruction, respectively. For timing evaluation,  the time residual $\Delta t = t_{\mathrm{rec}} - t_{\mathrm{true}}$ is computed for each true PE hit time $t_{\mathrm{true}}$, where $t_{\mathrm{rec}}$ is the closest reconstructed hit time. The mean and standard deviation of this residual distribution characterize the timing bias and resolution. The following analysis evaluates DFB and ResNet-50 reconstruction performance using EMC LT-UPE testing samples.

\begin{figure} [t]
    \centering
    \includegraphics[width=0.95\linewidth]{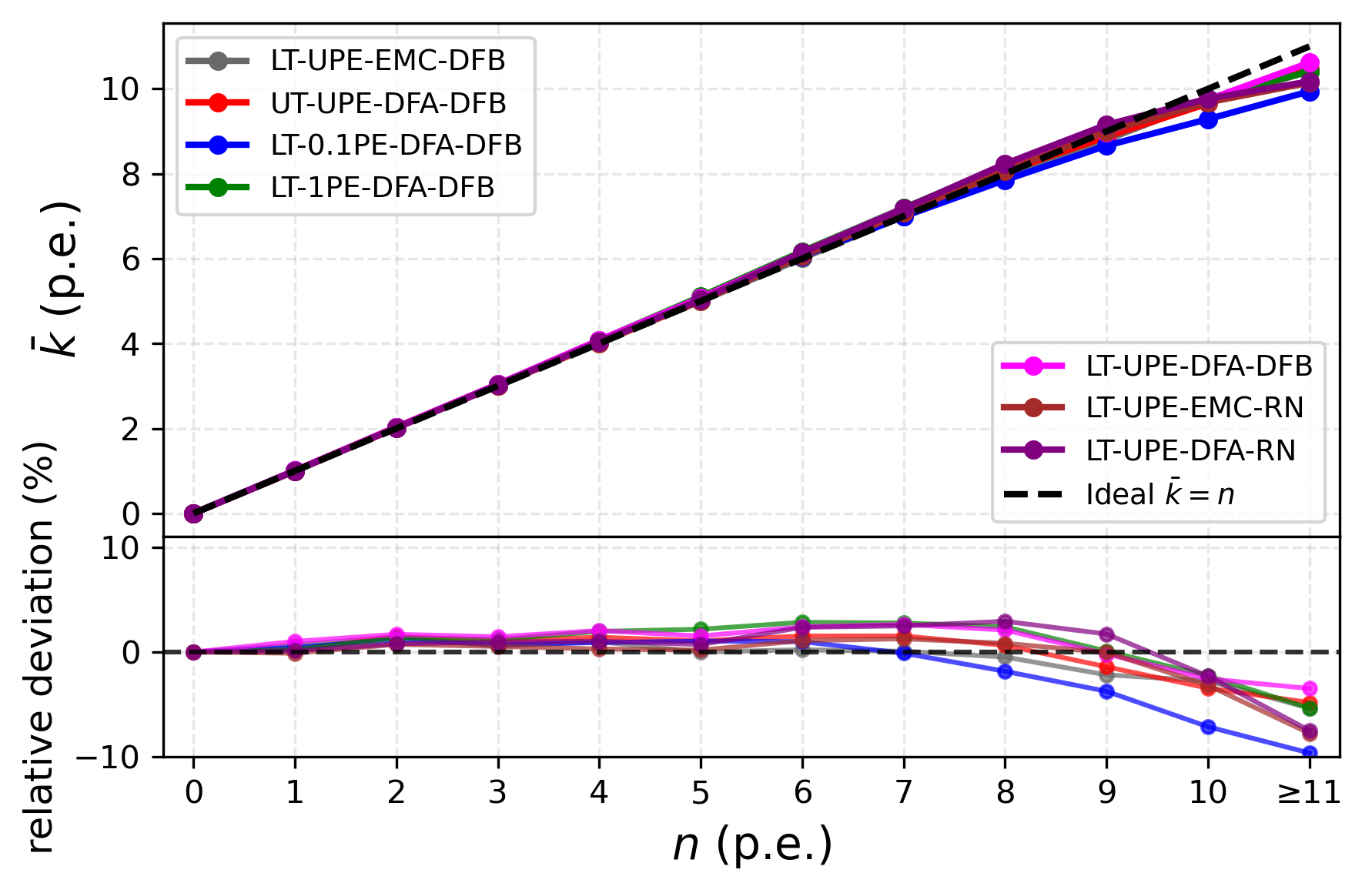}
    \includegraphics[width=0.95\linewidth]{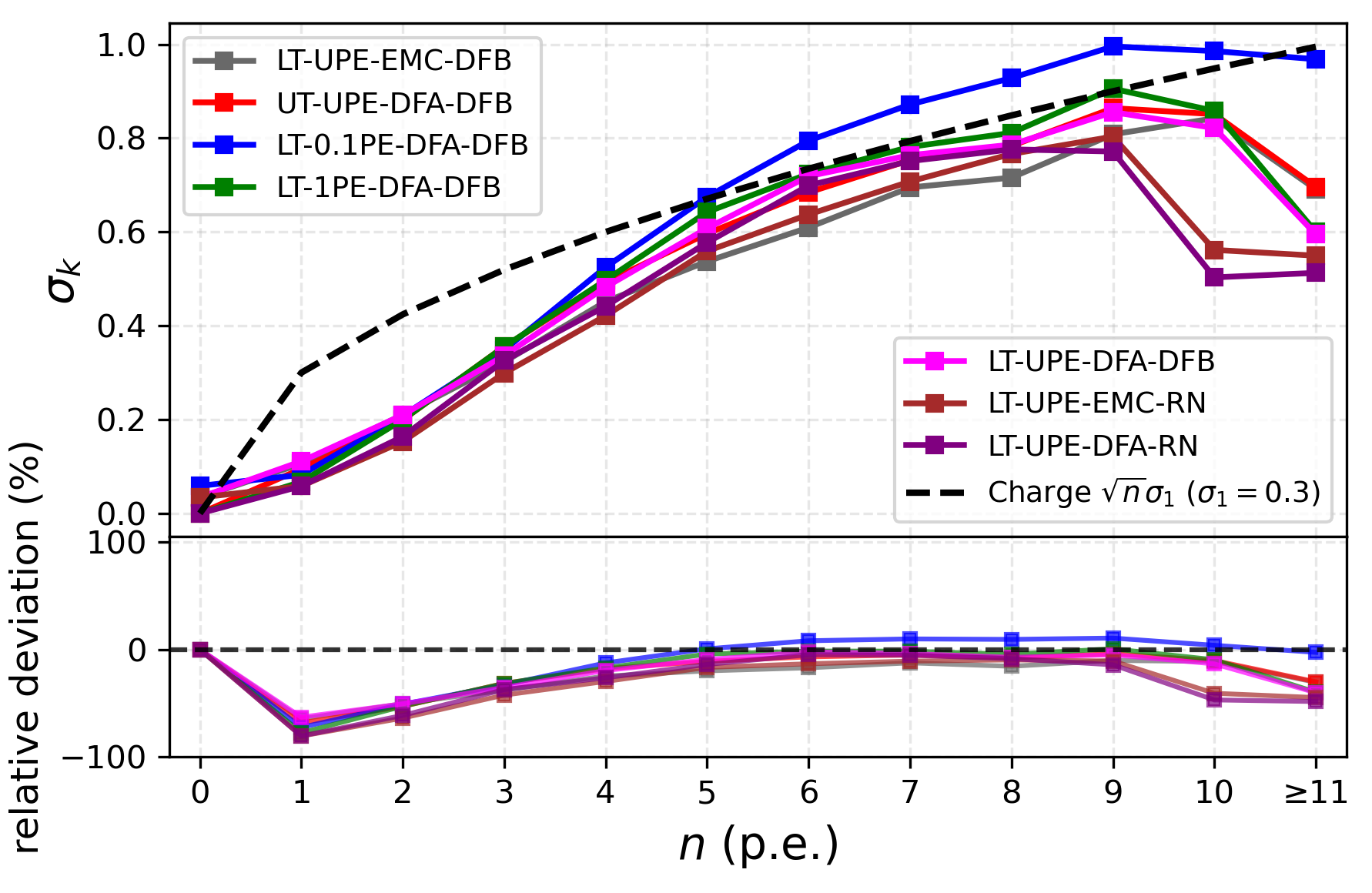}
    
    \caption{
        Comparison of nPE reconstruction performance across different models. The upper panel shows the reconstructed mean nPE $\bar{k}$ as a function of true nPE $n$ (solid lines) compared to the ideal linear relationship (dashed line). The relative deviation is defined as $(\bar{k} - n)/n$. The lower panel shows the reconstructed resolution $\sigma_{k}$ as a function of $n$ (solid lines) compared to the charge-based resolution $\sigma = \sqrt{n}\,\sigma_1$ (dashed line), where $\sigma_1 = 0.3$~p.e. is the sPE charge resolution. The relative deviation is defined as $(\sigma_{k} - \sigma)/\sigma$.
        For $n \leq 5$~p.e., all models exhibit an nPE bias below 2\% and outperform  the charge-based estimation in resolution.
    }
    \label{fig:NPErec}
\end{figure}
\begin{figure}
    \centering
    \includegraphics[width=0.95\linewidth]{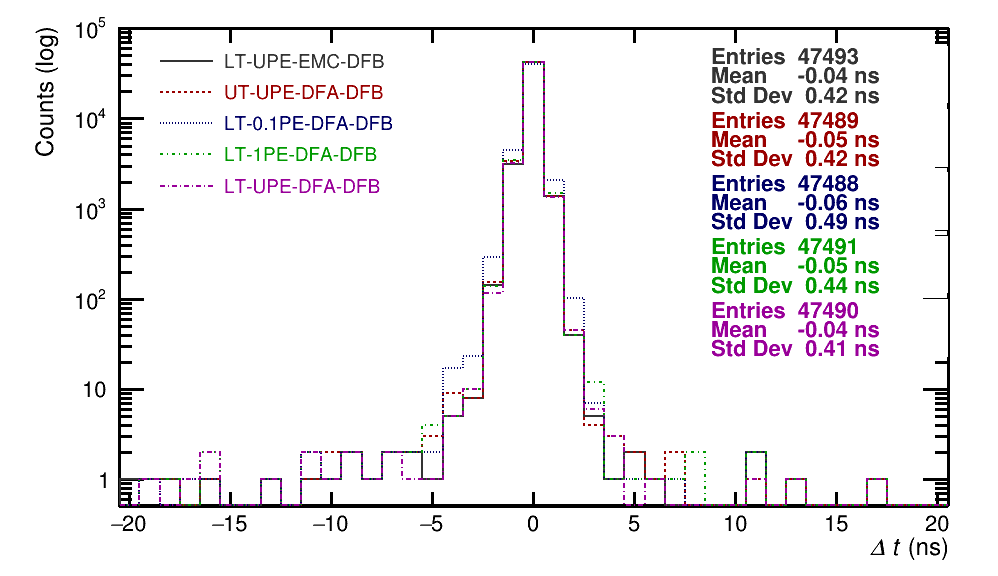}
    \caption{
        Distributions of PE timing residual $\Delta t$ for DFB models. All models achieve a timing resolution better than 0.5 ns, with LT-UPE-EMC-DFB and LT-UPE-DFA-DFB achieving the best performance.
    }
    \label{fig:T-sup}
\end{figure}

The nPE reconstruction performance is shown in Fig.~\ref{fig:NPErec}. In terms of bias, all models closely follow the ideal linear relationship $\bar{k} = n$, with deviations below 3\% for $n \leq 7$~p.e., indicating that charge nonlinearity from DFA has minimal impact on DFB performance. Regarding resolution, most models outperform the charge-based estimation, particularly for $n < 5$~p.e. where improvements are significant, for instance, achieving a standard deviation of approximately 0.1~p.e. at 1~p.e. This enhancement arises because PEs are temporally well-separated at low multiplicities, facilitating accurate identification by the models. For $n \geq 5$~p.e., increased degree of waveform overlap progressively obscures nPE-related features, causing resolution degradation across all models. Nevertheless, all models except LT-0.1PE-DFA-DFB maintain better resolution than the charge-based estimation. The performance degradation of LT-0.1PE-DFA-DFB stems from its DFA's training data lacking dense multi-PE samples, preventing effective modeling of features of overlapping waveforms. In the 5–8~p.e. range, LT-UPE-EMC-DFB and LT-UPE-EMC-RN achieve optimal performance; LT-UPE-DFA-RN, UT-UPE-DFA-DFB, LT-UPE-DFA-DFB, and LT-1PE-DFA-DFB show intermediate performance; while LT-0.1PE-DFA-DFB performs worst, consistent with the synthesis quality of their corresponding DFA models. These results demonstrate that both DFB and ResNet-50 can learn finer correlations between features of overlapping waveforms and nPE, thereby improving reconstruction accuracy, with DFB performance directly dependent on DFA's synthesis quality.

The PE timing performance of DFB models is shown in Fig.~\ref{fig:T-sup}. All models exhibit near-zero mean timing bias ($\approx$0 ns), confirming excellent unbiasedness in time reconstruction. In terms of timing resolution, LT-UPE-EMC-DFB, LT-UPE-DFA-DFB, and UT-UPE-DFA-DFB achieve the best performance with a standard deviations of approximately 0.42 ns; LT-1PE-DFA-DFB shows slightly degraded resolution ($\approx$0.44 ns); while LT-0.1PE-DFA-DFB exhibits the poorest resolution ($\approx$0.49 ns). This trend aligns with their nPE reconstruction performance.

In summary, for nPE reconstruction: (1) diffusion networks and ResNet-50 show comparable performance; (2) all models outperform the charge-based estimation in the low PE-multiplicity regime ($n < 5$~p.e.); (3) models trained on DFA-synthesized waveforms show slightly inferior performance compared to those trained on EMC waveforms, specifically, LT-UPE-DFA-DFB degrades by 7\% relative to LT-UPE-EMC-DFB, and LT-UPE-DFA-RN degrades by 3\% relative to LT-UPE-EMC-RN; and (4) among DFA-trained models, LT-UPE-DFA-DFB and UT-UPE-DFA-DFB achieve the best performance. For timing reconstruction, differences between DFB models are minor, with all showing timing resolution better than 0.5 ns. These results confirm that diffusion networks are well-suited for reconstructing PE sequences from PMT waveforms.

\section{Weakly Supervised Learning Approach}
Supervised learning methods require ground-truth PE information for training waveforms, which is difficult to obtain with high precision in practical applications. To address this limitation, this chapter proposes a weakly supervised learning approach based on BCDDPM described in Sec.~\ref{sec:BidiCondiDiffNetwork} for synergistic waveform simulation and reconstruction. The method initializes with coarse PE sequences and iteratively refines them using the BCDDPM until the reconstruction error of the PE sequences converges.  This weakly supervised mechanism only requires raw waveforms and coarse initial PE estimates to learn the mapping between waveform features and PE sequences. With appropriate training data selection, its reconstruction performance can approach that of fully supervised learning.

\subsection{Initial Labels}
\label{sec:xunfeng}
This subsection details the complete pipeline for extracting initial PE sequences from waveforms, which consists of three sequential steps: filtering, peak-finding, and selection. First, filtering suppresses electronic noise to minimize interference in subsequent peak detection. Next, peak-finding identifies candidate PEs. Finally, a dual-criteria selection based on amplitude and time interval removes residual false signals.

For the filtering stage, we systematically evaluated over ten filters using noise suppression effectiveness and waveform fidelity as evaluation metrics. The bilateral filter was ultimately selected due to its dual-domain smoothing mechanism, which operates in temporal and amplitude domains to preserve signal transition features while smoothing homogeneous regions. This characteristic is crucial for handling severely overlapping waveforms, effectively preventing PE information loss from over-smoothing. Optimized parameters are: neighborhood diameter $d = 9$ ns, amplitude-domain standard deviation $\sigma_{\mathrm{amp}} = 75$ ADC counts, and time-domain standard deviation $\sigma_{\mathrm{time}} = 75$ ns.

\begin{figure*}
    \centering
    \begin{subfigure}[b]{0.95\linewidth}
        \centering
        \includegraphics[width=\linewidth]{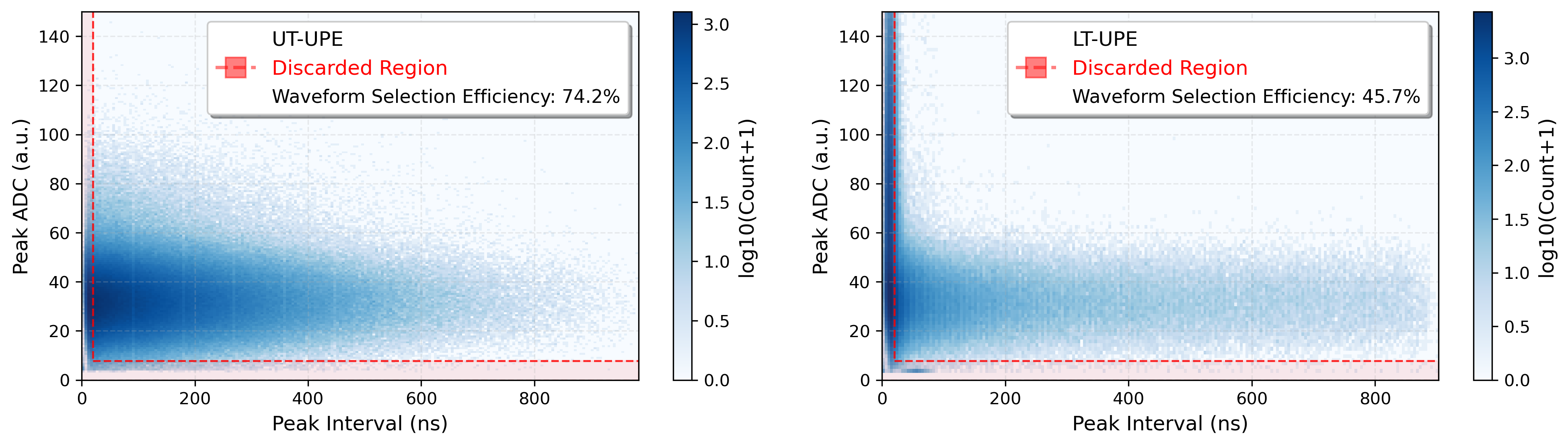}
        \caption{2D distribution of peak intervals versus peak amplitudes for adjacent predicted peak pairs in UT-UPE (left) and LT-UPE (right) samples.}
        \label{fig:FinkPeak_a}
    \end{subfigure}
    \vspace{1em} 
    \begin{subfigure}[b]{0.95\linewidth}
        \centering
        \includegraphics[width=\linewidth]{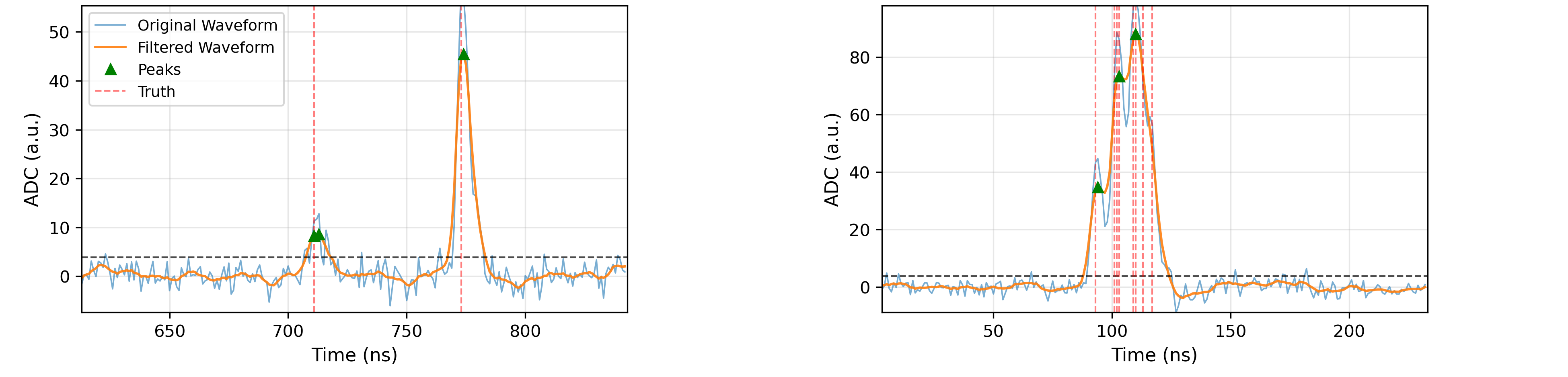}
        \caption{Typical examples of rejected waveforms: left panel shows noise-induced peaks in the low-amplitude, short-interval region; right panel shows non-sPE peaks caused by severe waveform overlap in the high-amplitude, short-interval region.}
        \label{fig:FinkPeak_b}
    \end{subfigure}
    \caption{Illustration of the selection criteria for reliable initial training data and labels: waveforms containing adjacent peaks with time separation less than 20 ns or with either peak amplitude below $2(\mu_N + 5\sigma_N)$ are discarded. 
    }
    \label{fig:FinkPeak}
\end{figure*}

After filtering, peaks exceeding the amplitude threshold $\mu_N + 5\sigma_N$ (where $\mu_N$ and $\sigma_N$ denote the mean and standard deviation of the filtered waveform baseline, respectively) are initially marked as PE candidates. For waveforms identified by the peak-finding as containing multiple PEs, we extract the time intervals of adjacent peaks and amplitudes of the subsequent peaks, producing 2D distributions of time interval versus peak amplitude for UT-UPE and LT-UPE samples, as shown in Fig.~\ref{fig:FinkPeak_a}. These distributions reveal two problematic regions: (1) low-amplitude, short-interval clusters from noise-induced peaks, and (2) high-amplitude, short-interval clusters due to severe waveform overlap (non-sPE peaks), as illustrated in Fig.~\ref{fig:FinkPeak_b}.

To address these issues, we apply a selection criterion: waveforms contain adjacent peaks separated by less than 20 ns or if either peak-amplitude below $2(\mu_N + 5\sigma_N)$ are discarded. The selection efficiencies are 74.2\% (UT-UPE), 99.2\% (LT-0.1PE), 84.3\% (LT-1PE), and 45.7\% (LT-UPE). Although this procedure sacrifices some severely overlapping waveforms, it provides higher-quality training data for the initial iteration of BCDDPM iterative training.

\subsection{Iterative Training}
\label{sec:looptrain}
Starting from the filtered waveforms $\mathbf{X}_{\text{init}}$ and their corresponding PE sequences $\mathbf{Y}_{\text{init}}$ obtained by the peak-finding algorithm, we construct an initial training set $\{\mathbf{X}_{\text{init}}, \mathbf{Y}_{\text{init}}\}$ to initialize BCDDPM training. This subsection details how iterative training enables joint optimization of waveform simulation and reconstruction.

Let the iteration index be $k = 0, 1, 2, \dots, K$. At iteration $k$, the real waveforms $\mathbf{X}^{(k)}$ and their reconstructed PE sequences $\mathbf{Y}^{(k)}$ serve as inputs to DFA, while the artificial PE sequences $\mathbf{Y}_{\text{gen}}$ and their associated synthetic waveforms $\mathbf{X}^{(k)}_{\text{gen}}$ are used as inputs to DFB. The time distribution of $\mathbf{Y}_{\text{gen}}$ is extracted with the peak-finding algorithm, and the nPE distribution of $\mathbf{Y}_{\text{gen}}$ is derived from the charge spectrum analysis. We initialize with $\mathbf{Y}^{(0)} = \mathbf{Y}_{\text{init}}$ and $\mathbf{X}^{(0)} = \mathbf{X}_{\text{init}}$. For $k \geq 1$, $\mathbf{X}^{(k)}$ is set to the unfiltered data $\mathbf{X}_{\text{orig}}$ to preserve the true proportion of overlapping waveforms. The data flow is illustrated in Fig.~\ref{fig:Diffusion-A-B}, and each iteration consists of four steps:

\begin{itemize}
    \item \textbf{Waveform generator optimization}: Train DFA conditioned on $\mathbf{Y}^{(k)}$ to learn the distribution of real waveforms $\mathbf{X}^{(k)}$, thereby improving its capability as a conditional waveform generator $p_\theta(\mathbf{x} \mid \mathbf{Y}^{(k)})$.
    
    \item \textbf{Simulated waveform update}: Use the optimized DFA to synthesize new waveforms $\mathbf{X}_{\text{gen}}^{(k)} \sim p_\theta(\mathbf{x} \mid \mathbf{Y}_{\text{gen}})$ based on the target PE sequence $\mathbf{Y}_{\text{gen}}$.
    
    \item \textbf{Waveform reconstruction optimization}: Train DFB conditioned on $\mathbf{X}_{\text{gen}}^{(k)}$ to synthesize the corresponding PE sequence $\mathbf{Y}^{(k)}_{\text{gen}}$, enhancing its performance as a PE sequence inference model $p_\theta(\mathbf{y} \mid \mathbf{X}_{\text{gen}}^{(k)})$.
    
    \item \textbf{PE sequence update}: Apply the improved DFB to reconstruct PE sequences from the original waveforms $\mathbf{X}_{\text{orig}}$, yielding refined estimates $\mathbf{Y}^{(k+1)} \sim p_\theta(\mathbf{y} \mid \mathbf{X}_{\text{orig}})$.
\end{itemize}

This cycle is repeated until predefined convergence criteria are met: (1) the charge linearity and resolution of waveforms synthesized by DFA across different time intervals of PEs no longer improve significantly; and (2) the nPE resolution of DFB stabilizes. Throughout this process, $\mathbf{X}_{\text{gen}}^{(k)}$ and $\mathbf{Y}^{(k)}$ are jointly refined through the bidirectional interaction between the diffusion models, leading to simultaneous improvements in both waveform simulation and reconstruction.

\subsection{Performance Analysis}
This section analyzes the waveform simulation and reconstruction performance of the BCDDPM after five iterations of weakly supervised training, using the UT-UPE, LT-1PE, LT-0.1PE, and LT-UPE EMC datasets described in Section~\ref{sec:data} as training data. Model naming conventions follow those established in Section~\ref{sec:sup-train}.

\begin{figure*} [t]
    \centering
    \begin{subfigure}[b]{0.32\linewidth}
        \centering
        \includegraphics[width=\linewidth, height=10cm]{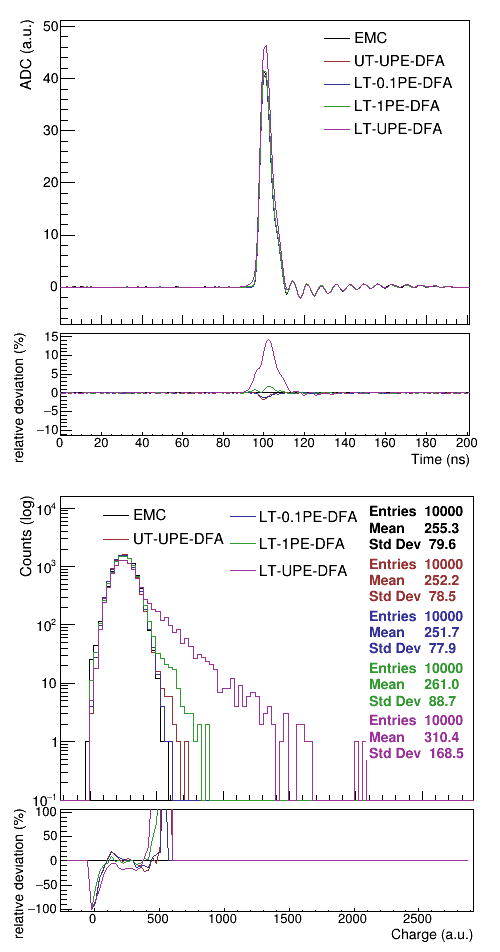}
        \caption{sPE samples}
        \label{fig:weak-DiffA_Check_a}
    \end{subfigure}
    \hfill
    \begin{subfigure}[b]{0.32\linewidth}
        \centering
        \includegraphics[width=\linewidth, height=10cm]{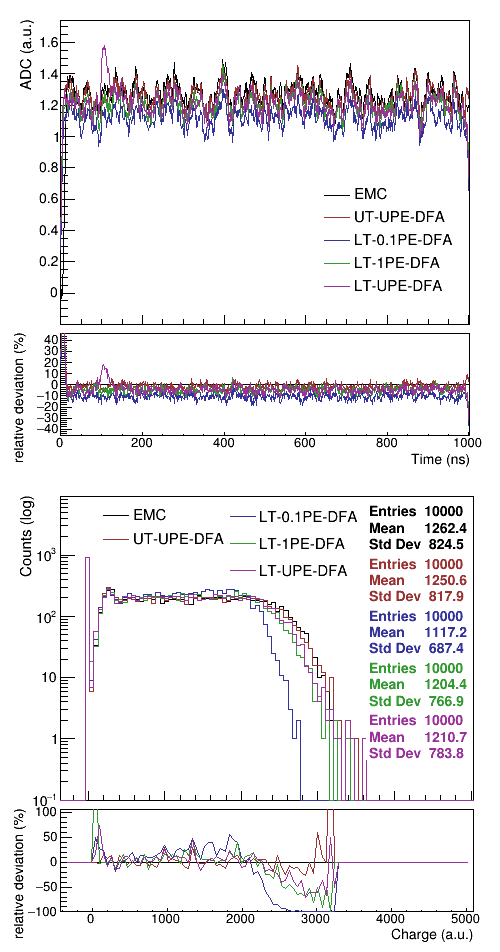}
        \caption{UT-UPE samples}
        \label{fig:weak-DiffA_Check_b}
    \end{subfigure}
    \hfill
    \begin{subfigure}[b]{0.32\linewidth}
        \centering
        \includegraphics[width=\linewidth, height=10cm]{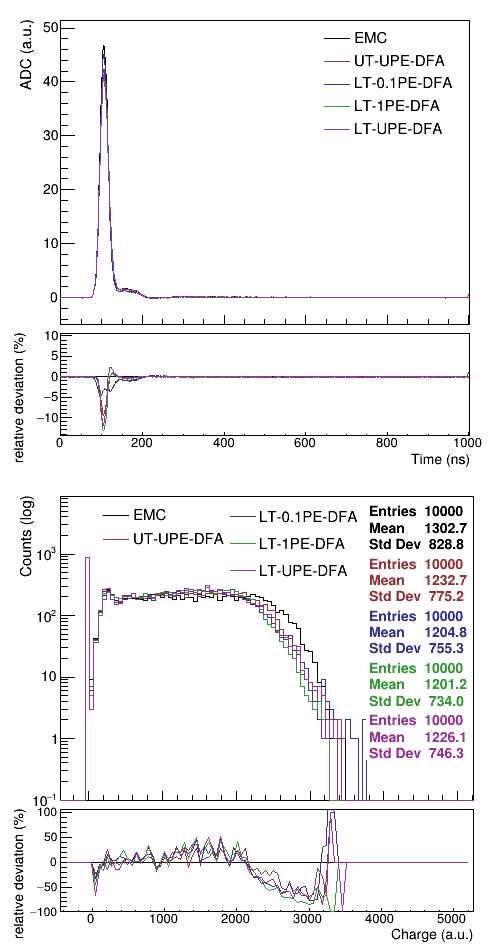}
        \caption{LT-UPE samples}
        \label{fig:weak-DiffA_Check_c}
    \end{subfigure}
    \caption{
        Comparison of the averaged waveforms and charge spectra between EMC and weakly supervised DFA models. The relative deviations are defined relative to EMC.
        (a) sPE samples: LT-UPE-DFA shows significant bias in both waveform shape and charge spectrum, while the other three models agree with EMC within a deviation of 2\%. 
        (b) UT-UPE samples: All four models exhibit uniform temporal distributions with overall waveform shapes consistent with EMC (deviations <5\%), except for LT-0.1PE-DFA, which shows $\sim$12\% waveform deviation. Comparisons of charge spectra reveal systematically lower charges in the high-nPE region for all models, most pronounced for LT-0.1PE-DFA. 
        (c) LT-UPE samples: UT-UPE-DFA, LT-UPE-DFA, and LT-1PE-DFA show $\sim$10\% waveform deviations, while LT-0.1PE-DFA maintains a maximum deviation of $\sim$4\%. Compared to its performance on UT-UPE samples, LT-0.1PE-DFA demonstrates reduced bias in the charge spectrum for LT-UPE samples.
    }
    \label{fig:weak-DiffA_Check}
\end{figure*}

\subsubsection{Waveform Simulation}
Fig.~\ref{fig:weak-DiffA_Check_a} compares the averaged sPE waveforms and charge spectra from EMC and weakly supervised DFA models. Relative to EMC, LT-UPE-DFA and LT-1PE-DFA exhibit averaged waveforms that are approximately 12\% and 1\% higher, respectively, with correspondingly elevated mean and standard deviation in the charge spectra. This bias stems from mislabeling of dense multi-PE as sPE waveforms in their training data. In contrast, LT-0.1PE-DFA and UT-UPE-DFA show averaged waveforms about 2\% lower than EMC, with correspondingly reduced mean and standard deviation of the charge spectra, closely matching their supervised counterparts, indicating minimal impact of weak supervision on sPE waveform simulation for these two models.

Fig.~\ref{fig:weak-DiffA_Check_b} shows the averaged waveforms and charge spectra for UT-UPE samples from EMC and weakly supervised DFA models. All DFA-synthesized waveforms exhibit uniform temporal distributions, demonstrating successful learning of the strong correlation between PE times and waveform peak positions under weak supervision. Compared to EMC, the waveform deviations are 0.8\% (UT-UPE-DFA), 12.0\% (LT-0.1PE-DFA), 5.0\% (LT-1PE-DFA), and 4.3\% (LT-UPE-DFA). Relative to supervised learning results, LT-0.1PE-DFA and LT-1PE-DFA show slightly reduced deviations, while UT-UPE-DFA and LT-UPE-DFA achieve performance comparable to their supervised versions. Both LT-0.1PE-DFA and LT-1PE-DFA synthesize waveforms with systematically lower charges in the high-PE region, a behavior consistent with their supervised counterparts, primarily due to insufficient multi-PE samples in their training data.

Fig.~\ref{fig:weak-DiffA_Check_c} presents the averaged waveform and charge spectra for LT-UPE samples from EMC and weakly supervised DFA models. Compared to EMC, the deviations of peak amplitude are 9.3\% (UT-UPE-DFA), 3.3\% (LT-0.1PE-DFA), 13.1\% (LT-1PE-DFA), and 11.4\% (LT-UPE-DFA), with all models showing charge deviations in the high-PE region. Notably, LT-0.1PE-DFA aligns more closely with EMC than its supervised counterpart, while the other models exhibit larger deviations compared to supervised learning. This further confirms that weak supervision has minimal impact on LT-0.1PE-DFA's performance.

\begin{figure*}[t]
     \centering

    \includegraphics[width=0.45\linewidth]{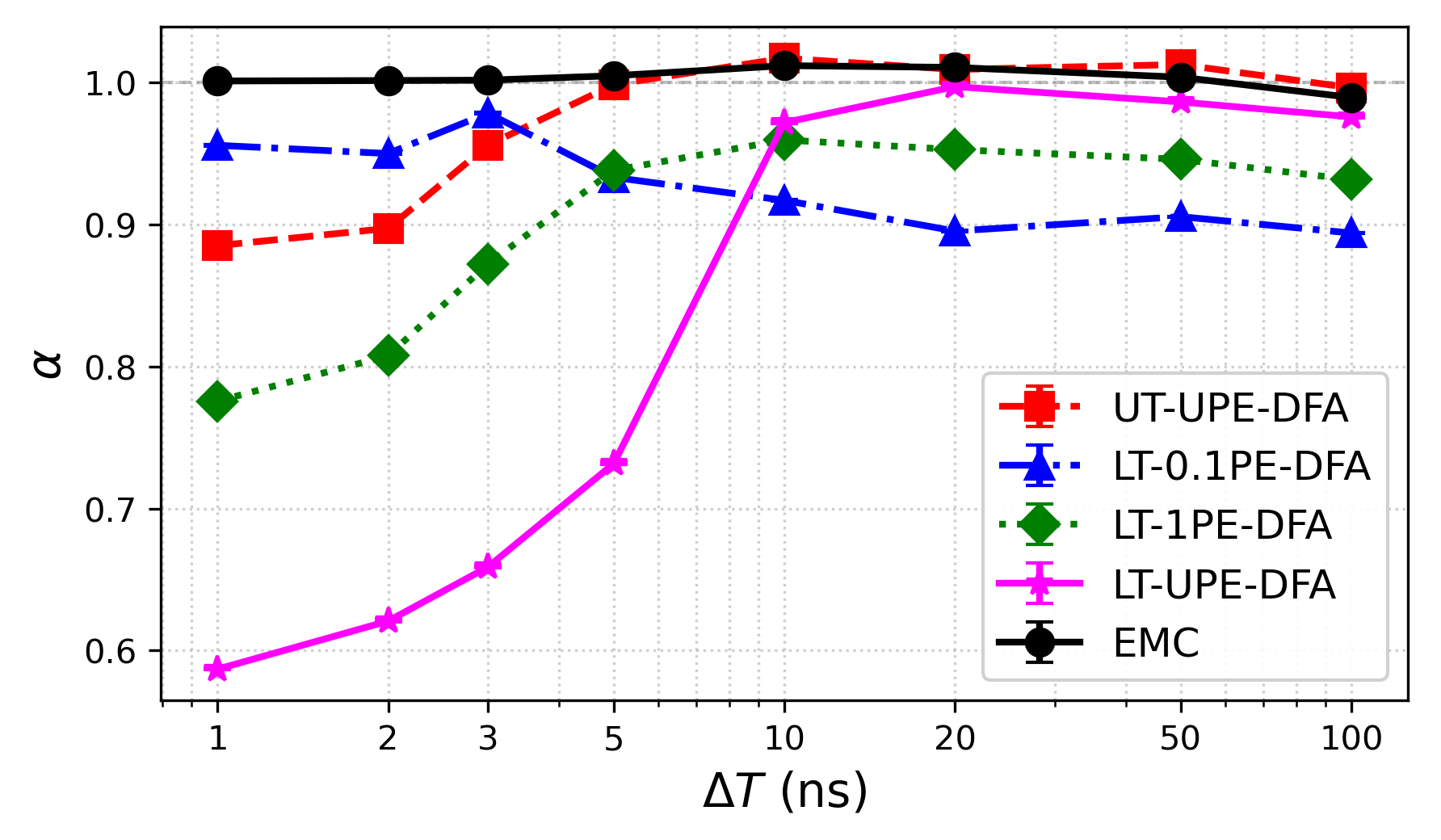}
    \includegraphics[width=0.45\linewidth]{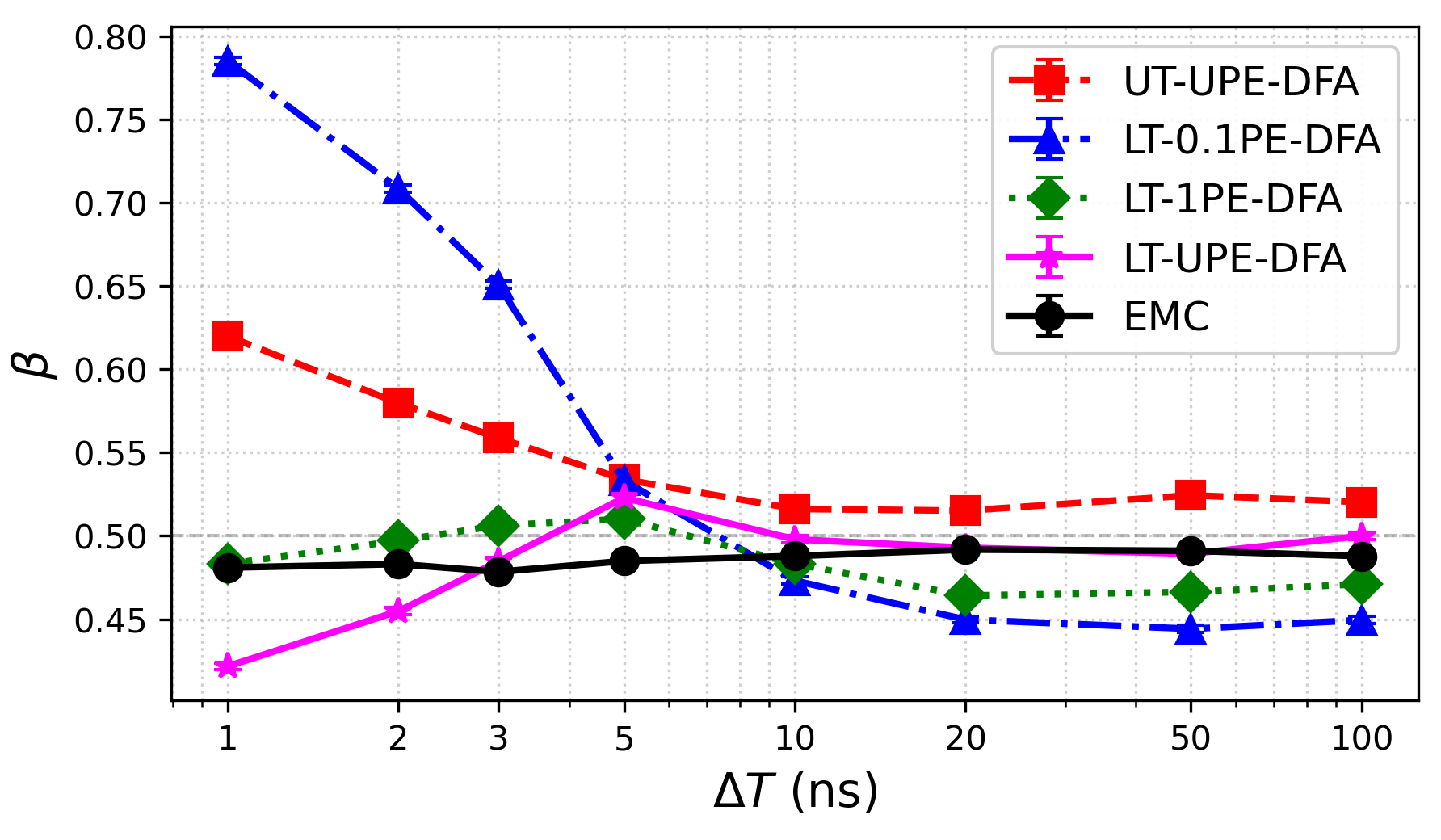}
  
    \caption{Dependence of the charge linearity coefficient $\alpha$ (left) and resolution exponent $\beta$ on $\Delta T$ (right) for EMC and weakly supervised DFA models. The EMC represent the ideal reference. For $\Delta T < 5~\textrm{ns}$, UT-UPE-DFA and LT-0.1PE-DFA exhibit optimal linearity, whereas LT-UPE-DFA shows significantly degraded performance. Among all models, LT-1PE-DFA yields a $\beta$ value closest to that of EMC, with deviation below 5\%. In contrast, LT-UPE-DFA and LT-0.1PE-DFA display resolution trends that are respectively lower and higher than that of  EMC, with LT-0.1PE-DFA showing a pronounced overestimation. As $\Delta T$ increases, the charge linearity of UT-UPE-DFA approaches that of EMC, while LT-0.1PE-DFA and LT-1PE-DFA deviate by approximately 5\% and 10\%,  respectively. Regarding resolution, the $\beta$ values of all models remain within 5\% of the EMC reference across the full $\Delta T$ range.}
    \label{fig:weak-QdTfit}
\end{figure*}

The fitted charge linearity coefficient $\alpha$ and resolution exponent $\beta$ as functions of $\Delta T$ are shown in Fig.~\ref{fig:weak-QdTfit}. In the severe overlap regime ($\Delta T < 5$ ns), LT-0.1PE-DFA and UT-UPE-DFA achieve the best charge linearity; LT-UPE-DFA shows significantly lower $\alpha$ than EMC; and LT-1PE-DFA exhibits slightly better linearity than LT-UPE-DFA. Regarding resolution, LT-1PE-DFA's $\beta$ is closest to EMC; UT-UPE-DFA and LT-UPE-DFA show elevated and reduced $\beta$, respectively; while LT-0.1PE-DFA exhibits substantially larger $\beta$.

As $\Delta T$ increases from 5 ns to 100 ns, UT-UPE-DFA's $\alpha$ gradually approaches the EMC value; LT-0.1PE-DFA and LT-1PE-DFA deviate from EMC by approximately 5\% and 10\%, respectively; and all models maintain $\beta$ within 5\% of the EMC value.

These results demonstrate that weakly supervised training with dense multi-PE samples (e.g., LT-UPE-DFA and LT-1PE-DFA) introduces significant charge nonlinearity in synthetic waveforms, whereas training with sparse multi-PE samples (e.g., LT-0.1PE-DFA and UT-UPE-DFA) yields waveform simulation performance closer to supervised learning. The next subsection analyzes how DFA performance impacts DFB reconstruction quality.

\begin{figure}
    \centering

    \includegraphics[width=0.95\linewidth]{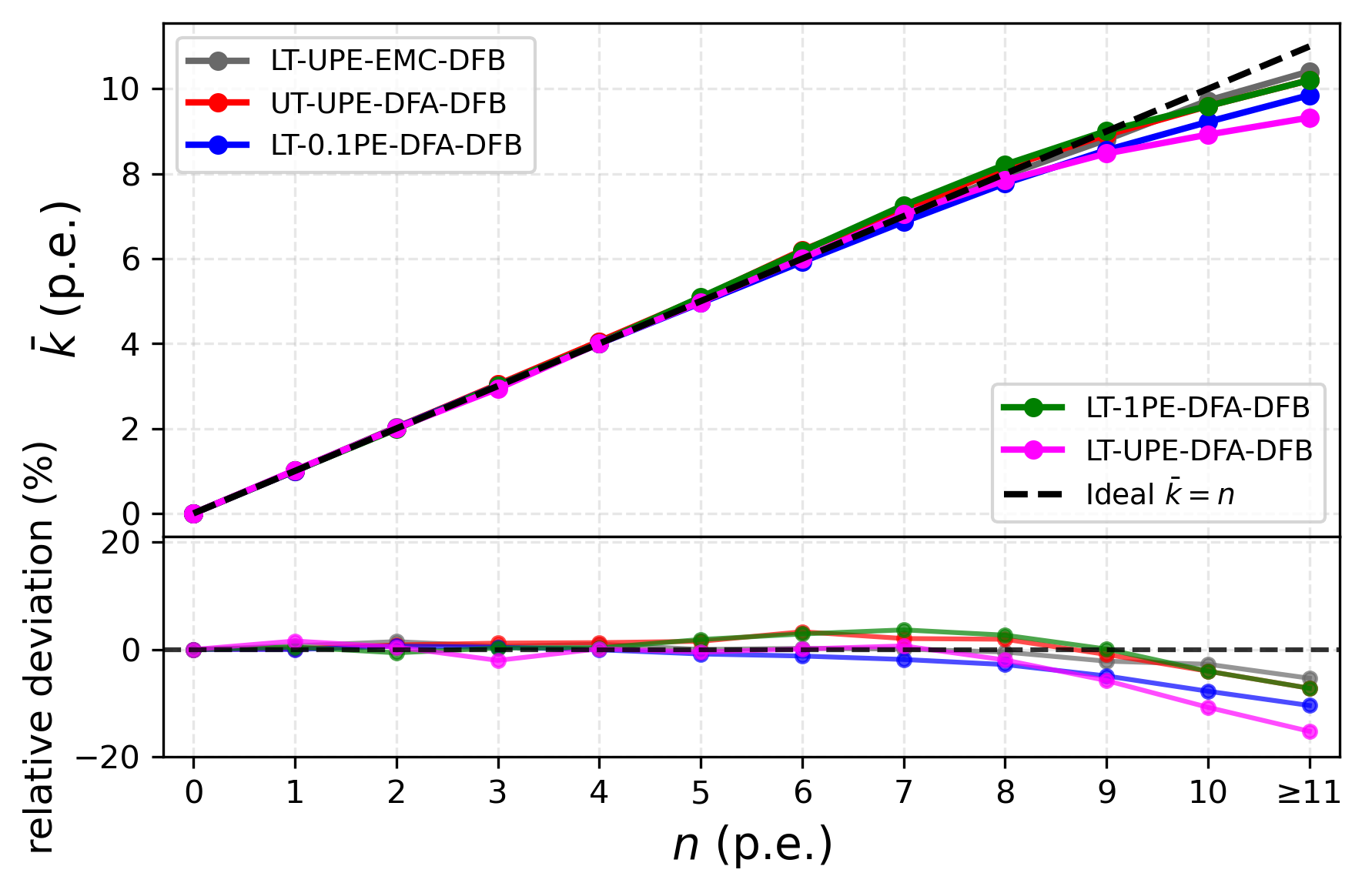}
    \includegraphics[width=0.95\linewidth]{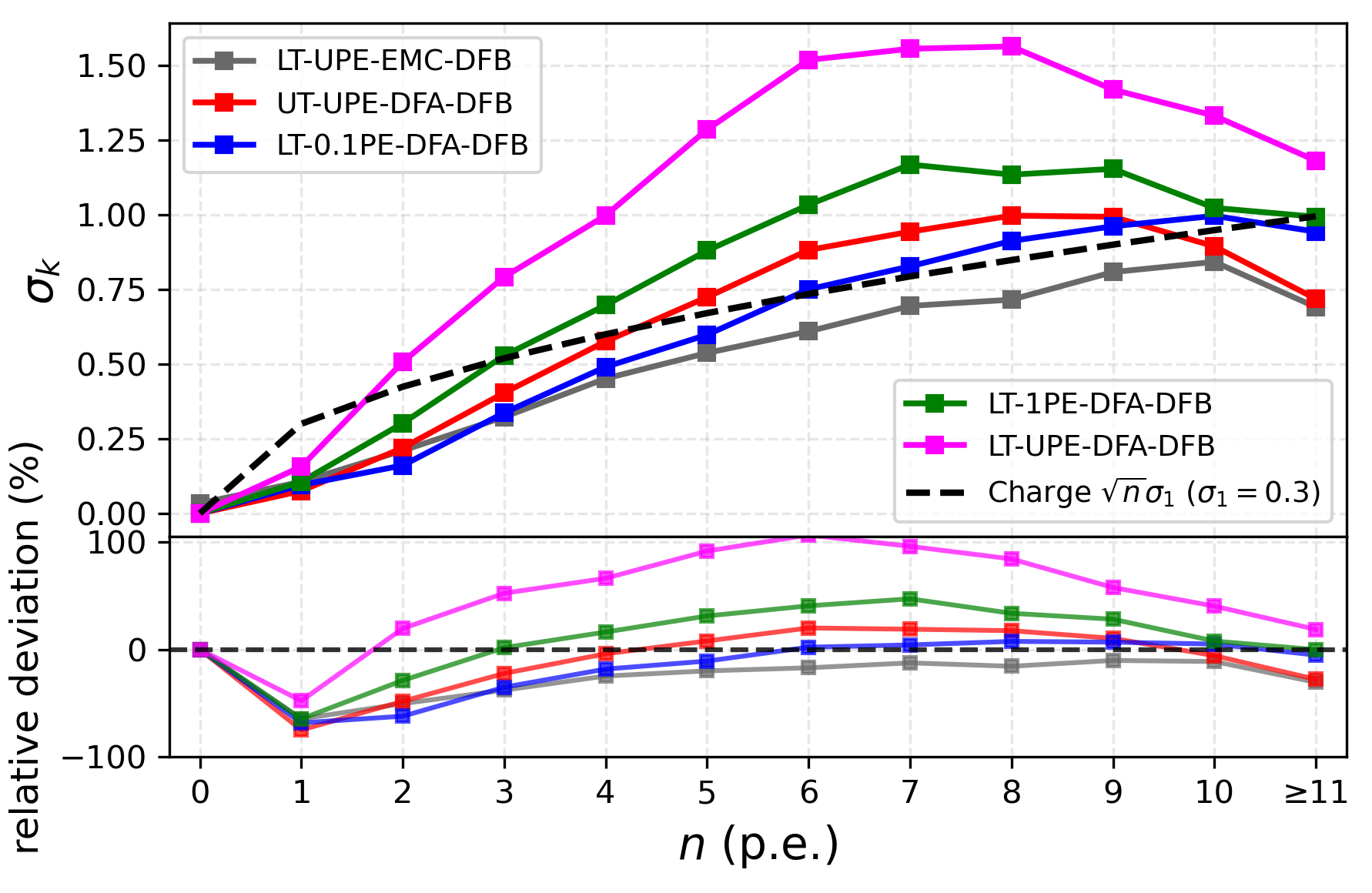}

    \caption{
        Comparison of the mean and standard deviation of reconstructed nPE for weakly supervised DFB models. The upper panel shows $\bar{k}$ as a function of true nPE $n$; the relative deviation is defined as $(\bar{k} - n)/n$, with the dashed black line indicating ideal linearity. The lower panel shows the reconstructed resolution $\sigma_{k}$ as a function of $n$ compared to the charge-based resolution $\sigma = \sqrt{n}\,\sigma_1$ (dashed black line), where $\sigma_1 = 0.3$~p.e. is the sPE charge resolution. The relative deviation is defined as $(\sigma_{k} - \sigma)/\sigma$. The degradation in nPE resolution correlates with the degree of waveform overlap in the training samples; notably, the weakly supervised LT-0.1PE-DFA-DFB model achieves resolution comparable to that of supervised baseline.
    }
    \label{fig:weak-NPErec}
\end{figure}

\begin{figure}
    \centering    
    \includegraphics[width=0.95\linewidth]{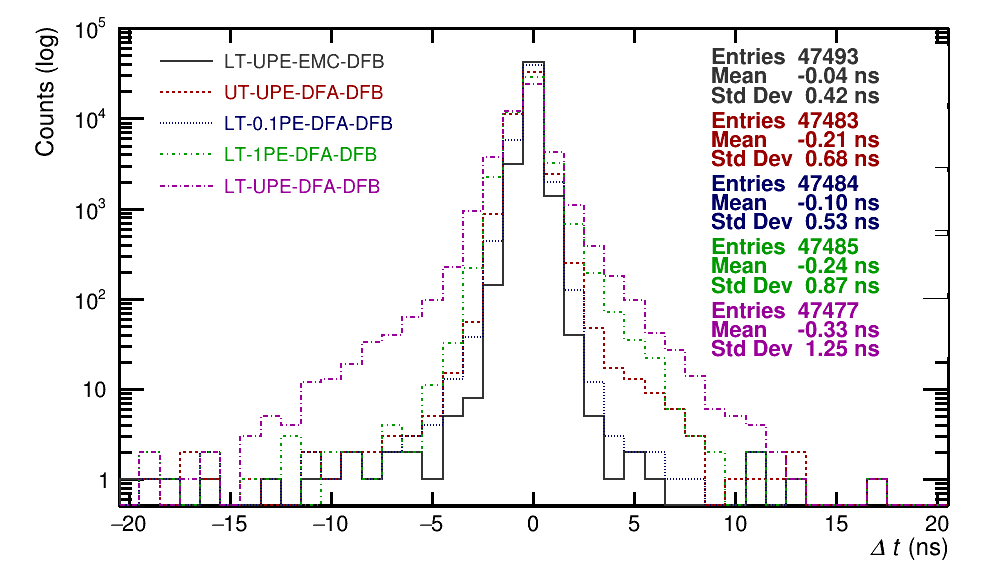}
    \caption{Distributions of PE timing residual $\Delta t$ for weakly supervised DFB models. The solid black histogram shows the timing residual from the supervised model LT-UPE-EMC-DFB, serving as the performance benchmark. All weakly supervised models reconstruct PE times earlier than the supervised model by $\sim$0.1–0.3~ns. Among them, LT-0.1PE-DFA-DFB achieves the best timing resolution, with a standard deviation of 0.53~ns, closest to the supervised performance.}
    \label{fig:weak-T-sup}
\end{figure}

\subsubsection{Waveform Reconstruction}
Fig.~\ref{fig:weak-NPErec} presents the nPE reconstruction performance of  weakly supervised DFB models, with the supervised model LT-UPE-EMC-DFB serving as baseline reference. In the region $n \leq 6$~p.e., all models exhibit a bias of less than 3\% in the reconstructed nPE, indicating that the charge nonlinearity present in DFA-synthesized waveforms has only a minor impact on the linearity of the nPE reconstruction by DFB. However, significant differences emerge in nPE resolution: only LT-0.1PE-DFA-DFB and UT-UPE-DFA-DFB achieve nPE resolution comparable to that of the supervised baseline, and notably outperform the charge-based estimation for $n \leq 4$~p.e. In contrast, LT-UPE-DFA-DFB and LT-1PE-DFA-DFB show substantially degraded resolution relative to the supervised model. This trend is consistent with the waveform simulation performance of their corresponding DFA models.

Fig.~\ref{fig:weak-T-sup} compares the PE timing performance of weakly supervised DFB models with the supervised baseline model LT-UPE-EMC-DFB. All weakly supervised models yield reconstructed times that are systematically earlier than the supervised reference by approximately 0.1–0.3~ns. In terms of timing resolution, LT-0.1PE-DFA-DFB performs best, with a standard deviation of 0.53~ns, closest to the supervised result. UT-UPE-DFA-DFB and LT-1PE-DFA-DFB achieve standard deviations of 0.68~ns and 0.87~ns, respectively, while UT-LPE-DFA-DFB exhibits the poorest resolution of 1.25~ns, significantly worse than the supervised model. This ranking in timing performance aligns with that observed in nPE reconstruction.

In summary, except for LT-0.1PE-DFA-DFB, all other weakly supervised DFB models exhibit degraded performance relative to supervised learning, with the extent of degradation positively correlated with the degree of waveform overlap in the training data. Severe overlap reduces the accuracy of the initial PE sequence labels, thereby introducing larger biases in waveform features during weakly supervised waveform simulation and ultimately degrading reconstruction accuracy. These results highlight that the accuracy of the initial PE sequence labels is a critical factor influencing the reconstruction performance of BCDDPM.

\section{Conclusions}
Improving the precision of PMT waveform reconstruction is crucial for enhancing detector energy resolution, which presents a longstanding challenge in particle physics experiments. To address the limitations of conventional methods and the heavy reliance of supervised learning on ground-truth labels, we propose an innovative weakly supervised approach based on a bidirectional conditional diffusion network framework. This framework consists of a waveform simulation network and a waveform reconstruction network, operating in a fully data-driven manner using only raw waveforms and coarse PE information, thereby significantly reducing the dependence on precise labels. Our study reveals that the PE distribution in the training data critically determines weakly supervised performance. Systematic evaluation shows that training on waveforms with an average intensity of $\sim$0.1~p.e. yields optimal results. Training with such data achieve a favorable balance: it contains sufficient sPE waveforms to accurately characterize the PMT response, includes mild overlap waveforms to capture realistic features of waveform overlap, and avoid severe  overlap waveforms that would otherwise introduce large errors in the initial PE sequences. Under this configuration, the weakly supervised method achieves an average normalized nPE resolution $E(\sigma_k / \sqrt{n})$ of 0.18~p.e. (99\% of the supervised learning value) over 1–5~p.e. and a timing resolution of 0.5~ns (80\% of the supervised learning value). 

This work provides an effective and practical approach for waveform simulation and reconstruction in particle physics experiments. Future efforts will focus on optimizing network architecture and extending this approach to broader physics analyses such as vertex and energy reconstruction.

\section*{Acknowledgements}
This work was partially supported by the National Natural Science Foundation of China (Grant No.12405231), the Characteristic Innovation Project for Regular Higher Education Institutions of Guangdong Provincial Department of Education (2024KTSCX044) and the Science Foundation of High-Level Talents of Wuyi University (2021AL027).

\begin{table*}
    \centering
\caption{Modified U-Net architecture. For Dense and Conv1D layers, the number in parentheses denotes the number of feature channels. For the down-sampling Strided-Conv1D layers, the numbers in parentheses denote the number of feature channels and the stride size. For the up-sampling UpSampling1D layers, the number in parentheses denotes the interpolation size.}
\label{tab:Conditional_U-Net}
    \renewcommand{\arraystretch}{1.6} 
    \setlength{\tabcolsep}{12pt} 
    \setcellgapes{4pt} 
    \makegapedcells
    \begin{tabular}{ccc}
    \hline\hline
    \textbf{Layer/Block} & \textbf{Input} & \textbf{Output shape} \\[4pt]
    \hline
    Input Block & 
      \makecell[c]{%
        $\left\{\begin{array}{c}
          \text{x:}(1000,1) \to \text{Conv1D(32)} \\[4pt]
          \text{y:}(1000,1) \to \text{Dense(32)} 
        \end{array}\right\}\text{Concat}$ \\[4pt]
        $\text{t(scalar)} \to \text{temb(256)}$ 
      } & 
      \makecell[c]{%
        $(1000, 64)$ \\[4pt]
        $(256)$
      } \\
    \hline
    \makecell[c]{% 
    $ \text{Down Block 1}$ \\[4pt]
    $ \text{(encoder)} $ }&
      \makecell[c]{%
        $\left\{\begin{array}{c}
          \text{Conv1D(64)} \\[4pt]
          \text{GroupNorm} \\[4pt]
          \text{Swish} \\[4pt]
          \text{Add temb(64)} \\[4pt]
          \text{Conv1D(64)} \\
        \end{array}\right\}\text{ResBlock(64)} \times2$ \\[4pt]
        $\text{Strided-Conv1D(64,2)}$
      } &
      \makecell[c]{%
        $(500, 64)$
      } \\[4pt]
    \hline
    Down Block 2 &
      \makecell[c]{%
        $\text{ResBlock(128)} \times2$ \\[4pt]
        $\text{Strided-Conv1D(128,2)}$
      } &
      \makecell[c]{%
        $(250, 128)$
      } \\
    \hline
    Down Block 3 &
      \makecell[c]{%
        $\text{ResBlock(256)} \times2$ \\[4pt]
        $\text{AttentionBlock(256)}$ \\[4pt]
        $\text{Strided-Conv1D(256,2)}$
      } &
      \makecell[c]{%
        $(125, 256)$
      } \\
    \hline
    Down Block 4 &
      \makecell[c]{%
        $\text{ResBlock(512)} \times2$ \\[4pt]
        $\text{AttentionBlock(512)}$ 
      } &
      \makecell[c]{%
        $(125, 512)$
      } \\
    \hline
    Middle block &
      \makecell[c]{% 
        $\text{ResBlock(512)}$ \\[4pt]
        $\text{AttentionBlock(512)}$ \\[4pt]
        $\text{ResBlock(512)}$ 
      } &
      \makecell[c]{%
        $(125, 512)$
      } \\
    \hline
    \makecell[c]{% 
    $ \text{Up block 1}$ \\[4pt]
    $ \text{(decoder)} $ }&
      \makecell[c]{%
        $\text{Concat(x+skip)}$ \\[4pt]
        $\text{ResBlock(512)} \times3$ \\[4pt]
        $\text{AttentionBlock(512)}$ \\[4pt]
        $\left\{\begin{array}{c}
          \text{UpSampling1D(2)} \\[4pt]
          \text{Conv1D(512)}
        \end{array}\right\}\text{UpSample(512)}$ 
      } &
      \makecell[c]{%
        $(250, 512)$
      } \\
    \hline
    Up Block 2 &
      \makecell[c]{%
        $\text{Concat(x+skip)}$ \\[4pt]
        $\text{ResBlock(256)} \times3$ \\[4pt]
        $\text{AttentionBlock(256)}$ \\[4pt]
        $\text{UpSample(256)}$
      } &
      \makecell[c]{%
        $(500, 256)$
      } \\
    \hline
    Up Block 3 &
      \makecell[c]{%
        $\text{Concat(x+skip)}$ \\[4pt]
        $\text{ResBlock(128)} \times3$ \\[4pt]
        $\text{UpSample(128)}$
      } &
      \makecell[c]{%
        $(1000, 128)$
      } \\
    \hline
    Up Block 4 &
      \makecell[c]{%
        $\text{Concat(x+skip)}$ \\[4pt]
        $\text{ResBlock(64)} \times3$ 
      } &
      \makecell[c]{%
        $(1000, 64)$
      } \\
    \hline
    End block &
      \makecell[c]{% 
        $\text{GroupNorm}$ \\[4pt]
        $\text{Swish}$ \\[4pt]
        $\text{Conv1D(1)}$ 
      } &
      \makecell[c]{%
        $(1000, 1)$
      } \\
    \hline
    \hline
    \end{tabular}   
\end{table*}

\begin{table*}[ht]
    \centering
\caption{Hyperparameters of BCDDPM. The hyperparameter combination for the diffusion process, network architecture, and training parameters were gradually determined through supervised grid search until the training performance stabilized.}
\label{tab:DDPM_hyperparams} 
\renewcommand{\arraystretch}{1.2} 
\begin{tabular}{p{5.2cm} p{6.0cm}}
    \toprule
    \textbf{Parameter} & \textbf{Value} \\
    \midrule
    Diffusion Utility & Gaussian Diffusion \\
    Timesteps (T) & 200 \\
    Beta Schedule for Diffusion-A & 
        \makecell[l]{%
        $\left\{\begin{array}{l}
          \beta_{\text{start}}=1\times10^{-4},\ \beta_{\text{end}}=1\times10^{-1} \\
          \text{Linear Schedule}
        \end{array}\right.$ 
      } \\
    Beta Schedule for Diffusion-B & 
        \makecell[l]{%
        $\left\{\begin{array}{l}
          \beta_{\text{start}}=1\times10^{-3},\ \beta_{\text{end}}=1\times10^{-2} \\
          \text{Linear Schedule}
        \end{array}\right.$ 
      } \\
    Clip Range & $\text{clip-min}=-100.0,\ \text{clip-max}=1000.0$\\
    \midrule
    Input Shape & $(1000,1)$  \\
    Channel Widths &  [64, 128, 256, 512]\\
    Attention Locations & [False, False, True, True]\\
    Time MLP units & 256 \\
    Normalization & Group Normalization (groups = 8) \\
    Activation & Swish  \\
    \midrule
    Optimizer & Adam \\
    Optimizer Parameters & 
        \makecell[l]{%
            $\left\{\begin{array}{l}
              \text{Betas}: \beta_1=0.9,\ \beta_2=0.999 \\
              \text{Weight decay}: None \\
              \text{Learning rate}: 2\times10^{-4}
            \end{array}\right.$ 
          } \\
    Loss Function & Mean Squared Error(MSE) \\
    Batch Size  & 256 \\
    Training Strategy & Mirrored Strategy(multi-GPU) \\
    Training Epochs & 50 \\
    \bottomrule
    \end{tabular}
\end{table*}

\begin{table*}[h]
    \centering
\caption{Hyperparameters of ResNet50. The hyperparameters in the table were selected through repeated experimentation and tuning on the validation set during the supervised training of ResNet50. This combination provides robust and reproducible training behavior.}
\label{tab:ResNet50_hyperparams}
\renewcommand{\arraystretch}{1.2} % 增加行高
\begin{tabular}{p{5.2cm} p{6.0cm}}
    \toprule
    \textbf{Parameter} & \textbf{Value} \\
    \midrule
    Optimizer & AdamW \\
    Optimizer Parameters & 
        \makecell[l]{%
            $\left\{\begin{array}{l}
              \text{Betas}: \beta_1=0.9,\ \beta_2=0.999 \\
              \text{Weight decay}: 10^{-4} \\
            \end{array}\right.$ 
          } \\
    Learning Rate Strategy & ReduceLROnPlateau \\
    Learning Rate Parameters & 
        \makecell[l]{%
            $\left\{\begin{array}{l}
              \text{Learning rate}: 1\times10^{-3} \\
              \text{Factor}: 0.5 \\
              \text{Patience}: 5 \\
              \text{Minimum learning rate}: 1\times10^{-6} \\
            \end{array}\right.$ 
          } \\
    Activation & ReLU  \\
    Loss Function & SparseCategoricalCrossentropy \\
    Training Epochs & 100 \\
    Batch Size  & 256 \\
    Gradient Clipping & 0.1 \\
    \bottomrule
    \end{tabular}
\end{table*}

\end{document}